\documentclass[twocolumn,floatfix]{aastex631}

\newcommand{\ra}[1]{\renewcommand{\arraystretch}{#1}}

\usepackage{booktabs}
\usepackage{float}
\usepackage{gensymb}
\usepackage{array}
\usepackage{multirow}
\usepackage{graphicx}

\shorttitle{Warhol Stars}
\shortauthors{Williams et al.}

\graphicspath{{./}{figures/}}

\begin{document}

\title{JWST's PEARLS: Temperatures of Nine Highly Magnified Stars in a Galaxy at Redshift 0.94 and Simulated Stellar Population Dependence on Stellar Metallicity and the Initial Mass Function}

\date{2 July 2025}

\author[0000-0002-1681-0767]{Hayley Williams}
\affiliation{School of Physics and Astronomy, University of Minnesota 
116 Church Street SE, Minneapolis, MN 55455, USA 
}
\affiliation{School of Earth and Space Exploration, Arizona State University, Tempe, AZ 85287-6004, USA}

\author[0000-0003-3142-997X]{Patrick L. Kelly}
\affiliation{School of Physics and Astronomy, University of Minnesota 
116 Church Street SE, Minneapolis, MN 55455, USA 
}
\author[0000-0001-8156-6281]{Rogier A. Windhorst}
\affiliation{School of Earth and Space Exploration, Arizona State University, Tempe, AZ 85287-6004, USA}

\author[0000-0003-3460-0103]{Alexei V. Filippenko}
\affiliation{Department of Astronomy, University of California, Berkeley, CA 94720-3411, USA}

\author[0000-0003-1276-1248]{Amruth Alfred}
\affiliation{Department of Physics, The University of Hong Kong, Pokfulam Road, Hong Kong}

\author[0000-0002-8785-8979]{Tom Broadhurst}
\affiliation{Donostia International Physics Center, DIPC, Basque Country, San Sebasti\'an, 20018, Spain}
\affiliation{Department of Physics, University of Basque Country UPV/EHU, Bilbao, Spain}
\affiliation{Ikerbasque, Basque Foundation for Science, Bilbao, Spain}

\author[0000-0003-1060-0723]{Wenlei Chen}
\affiliation{Department of Physics, Oklahoma State University, 145 Physical Sciences Bldg, Stillwater, OK 74078, USA}
\author[0000-0003-1949-7638]{Christopher J.\ Conselice} 
\affiliation{Jodrell Bank Centre for Astrophysics, Alan Turing Building,
University of Manchester, Oxford Road, Manchester M13 9PL, UK}

\author[0000-0003-3329-1337]{Seth H. Cohen} 
\affiliation{School of Earth and Space Exploration, Arizona State University,
Tempe, AZ 85287-1404, USA}

\author[0000-0001-9065-3926]{Jos\'e M. Diego}
\affiliation{IFCA, Instituto de F\'isica de Cantabria (UC-CSIC), Av. de Los Castros s/n, 39005 Santander, Spain}

\author[0000-0002-4884-6756]{Benne W. Holwerda}
\affiliation{Department of Physics, University of Louisville, Natural Science Building 102, 40292 KY Louisville, USA}

\author[0000-0002-6610-2048]{Anton M. Koekemoer}
\affiliation{Space Telescope Science Institute, 3700 San Martin Drive,
Baltimore, MD 21218, USA}

\author[0000-0002-4490-7304]{Sung Kei Li}
\affiliation{Department of Physics, The University of Hong Kong, Pokfulam Road, Hong Kong}

\author[0000-0002-7876-4321]{Ashish Kumar Meena}
\affiliation{Physics Department, Ben-Gurion University of the Negev, P.O. Box 653, Be’er-Sheva 84105, Israel}

\author[0000-0003-0942-817X]{Jos\'e M. Palencia}
\affiliation{IFCA, Instituto de F\'isica de Cantabria (UC-CSIC), Av. de Los Castros s/n, 39005 Santander, Spain}

\author[0000-0003-4223-7324]{Massimo Ricotti}
\affiliation{Department of Astronomy, University of Maryland, College Park, MD 20742, USA}

\author[0000-0002-5404-1372]{Clayton D. Robertson}
\affiliation{Department of Physics, University of Louisville, Natural Science Building 102, Louisville, KY 40292, USA}

\author[0000-0001-7957-6202]{Bangzheng Sun}
\affiliation{Department of Physics and Astronomy, University of Missouri - Columbia, Columbia, MO 65201, USA}


\author[0000-0001-7592-7714]{Haojing Yan}
\affiliation{Department of Physics and Astronomy, University of Missouri - Columbia, Columbia, MO 65201, USA}

\author[0000-0002-0350-4488]{Adi Zitrin}
\affiliation{Physics Department, Ben-Gurion University of the Negev, P.O. Box 653, Be’er-Sheva 84105, Israel}

\begin{abstract}
We present stellar atmosphere modeling of {\it JWST} NIRCam photometry of nine highly magnified individual stars in a single galaxy at redshift $z=0.94$ known as the Warhol arc, which is strongly lensed by the galaxy cluster MACS\,J0416. Seven of these transients were identified by Yan et al. (2023). The nine sources are likely red supergiants with temperatures $T_{\rm eff}\approx 4000$\,K. We present new long-slit spectroscopy of the Warhol arc acquired with Keck-I telescope and the Large Binocular Telescope, and use these data to help constrain the arc's oxygen abundance to be $12+\log({\rm O}/{\rm H})=8.45\pm0.08$. A microlensing simulation is performed on synthetic stellar populations using a range of stellar metallicities and initial mass function (IMF) slopes. The temperature distribution of the simulated detectable stars is sensitive to the choice of stellar metallicity, and setting the stellar metallicity equal to the arc's nebular metallicity ($\log(Z_*/Z_\odot)=-0.24$) produces a simulated temperature distribution that is consistent with the observations, while lower stellar metallicities ($\log(Z_*/Z_\odot)<-0.75$) produce simulated temperatures that are inconsistent with the observations. The expected detection rate is strongly anticorrelated with the IMF slope for $\alpha>1.2$. For the canonical IMF slope $\alpha=2.35$, the simulation yields expected transient detection rates that agree with the observed detection rates in the {\it HST} Flashlights filters, but overpredicts the detection rate by a factor of $\sim 3$--12 ($<2\sigma$ tension) in the {\it JWST} filters. The simulated detection rate is sensitive to the choice of stellar metallicity, with lower metallicities ($\log(Z_*/Z_\odot)<-0.75$) yielding a significantly lower simulated detection rate that further reduces the modest tension with the observations. 
\end{abstract}


\section{Introduction}
Galaxy-cluster gravitational lenses are powerful tools for studying magnified background galaxies with much greater resolution and sensitivity than is otherwise possible. Lensing magnification is greatest for small sources that are located adjacent to the critical curve of a cluster. Individual stars at cosmological distances can be detected and analyzed during transient events, where the magnification of a star in a background galaxy lying close to the cluster's critical curve is temporarily greatly boosted by microlensing from an intracluster star or other compact object. 

\begin{figure*}[ht]
    \includegraphics[width=\linewidth]{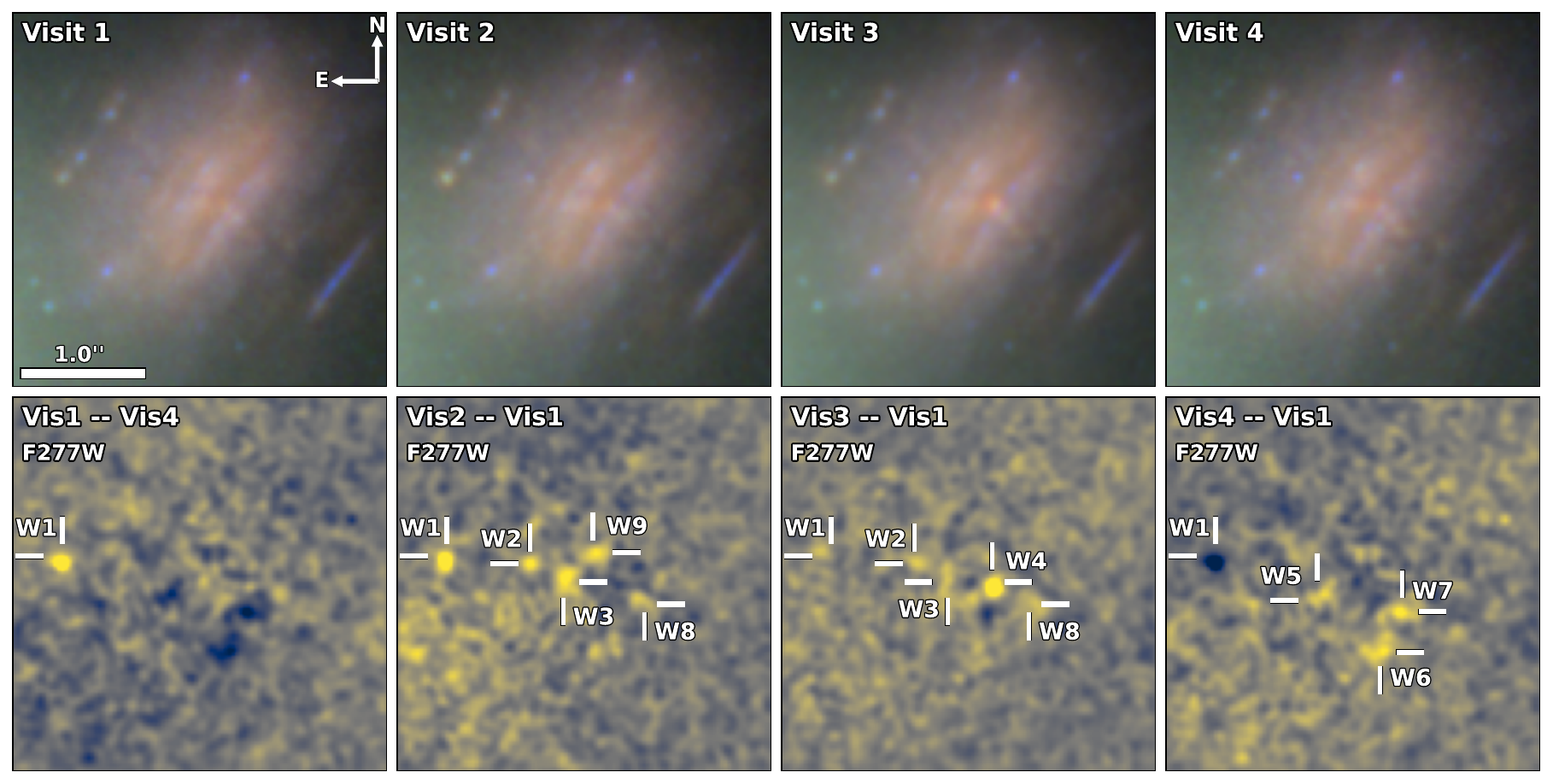}
    \caption{{\it Top:} NIRCam RGB color images of the Warhol arc in all four visits. Blue, F090W + F115W + F150W; Green, F200W + F277W; Red, F356W + F410M + F444W. {\it Bottom:} Difference images between the different visits in the F277W filter, with the 9 transients identified. Sources W1$-$W7 were originally identified by \cite{Yan_2023}. To aid in visual identification of the transients, these difference images have been convolved with a 1-pixel Gaussian kernel. {The orientation of these images is North pointing upward and East pointing to the left.}}
    \label{fig:warhol_cutouts}
\end{figure*}

The first detection of one of these events was the discovery in {\it Hubble Space Telescope (HST)} imaging of ``Icarus," a blue supergiant star in a spiral galaxy at redshift $z=1.49$ whose magnification was boosted up to $\mu \approx 2000$ due to microlensing by a compact object in the MACS\,J1149.5+2223 cluster \citep{Kelly_2018}. Similar events have also been discovered in {\it HST} imaging of other lensed galaxies at $z\approx 1$  \citep[e.g.,][]{Rodney_2018,Chen_2019,Kaurov_2019,Kelly_2022}. {\it HST} imaging of the WHL0137–08 cluster field led to the discovery of ``Earendel," a highly magnified object at $z=6.2$ \citep{Welch_2022}, but flux variation from microlensing has not yet been detected for this event.

{\it JWST}'s unique sensitivity in the infrared (IR) is improving our ability to detect and characterize microlensing transient events in cluster fields. With two {\it JWST} NIRCam visits to the Abell 370 galaxy-cluster lens separated by $\sim 1$\,yr, \cite{Fudamoto_2025} identified more than 40 magnified stars in a single lensed caustic-crossing galaxy known as the ``Dragon arc" at $z=0.725$. \cite{Welch_2022b} followed up Earendel with {\it JWST} NIRCam imaging and constrained the lower limit on the object's magnification to $\mu>4000$, and placed constraints on the object's temperature and source-plane size (see \citealt{Ji_2025} for a discussion on possible caveats regarding these constraints). 

Detections of lensed stars at cosmological distances present an exciting opportunity for probing a wide range of science questions. With deep multiband imaging, we can constrain the physical properties of individual massive stars in distant galaxies to better understand their evolution. Lensed stars can be used to measure the distribution of dark matter subhaloes \citep{WilliamsLLR_2024}, constrain the stellar initial mass function (IMF) at intermediate to high redshifts \citep{Li_2025,Palencia_2025}, and potentially discriminate between dark matter models \citep{Amruth_2023,Diego_2023,Broadhurst_2025}.

Prime Extragalactic Areas for Reionization and Lensing Science \citep[PEARLS; ][]{Windhorst_2023} is a {\it JWST} program with significant time devoted to NIRCam monitoring of the MACS\,J0416.1-2403 lensing cluster (hereafter M0416; $z=0.397$), which is one of the Hubble Frontier Fields \citep[HFFs;][]{Lotz_2017}. The PEARLS program obtained three epochs of NIRCam imaging of M0416, and an additional NIRCam epoch was obtained by another {\it JWST} program, the CAnadian NIRISS Unbiased Cluster Survey \citep[CANUCS; ][]{Willott_2022,Sarrouh_2025}. With four epochs of NIRCam imaging spanning 126 days, M0416 is a prime field for detecting and studying transient events.

A transient search was performed by \cite{Yan_2023} (hereafter Y23) using the four-epoch NIRCam data of M0416. 14 transients were detected, 12 of which were likely highly magnified individual stars. Seven of these transients were found in a lensed background galaxy at $z=0.94$, known as the ``Warhol'' arc. The eight-band NIRCam photometry spanning a wavelength range of 0.9\,$\mu$m to 4.4\,$\mu$m  enabled the construction of the spectral energy distribution (SED) of each transient source from 0.5\,$\mu$m to 2.2\,$\mu$m in the rest frame. 

Here we present the SED analysis and characterization of the transient sources found by the PEARLS program in the Warhol arc at $z = 0.94$. We use Bayesian inference to fit a stellar model to the SED for each source and constrain their temperatures, surface gravities, and magnitudes of extinction due to dust. We present new near-IR spectroscopy of the Warhol arc obtained with the combined 11.8\,m Large Binocular Telescope (LBT) and the 10\,m Keck-I telescope, and we use measurements of Balmer flux ratios to constrain dust attenuation in the arc. The observed detection rate and inferred stellar properties of the stars in the Warhol arc are compared with a stellar population synthesis simulation of expected microlensing events, and we test the impact of the slope of the IMF on the simulated population and expected detection rates. 

This paper is organized as follows. Section \ref{sec:observations} describes the {\it JWST}, LBT, and Keck observations and data reduction. Our transient search method is described in Section \ref{sec:transient_search}. In Section \ref{sec:measurements} we detail our methods for measuring the photometry of the transients from the NIRCam data and for measuring the emission-line fluxes from the LBT and Keck spectroscopy. We describe our stellar SED fitting of the Warhol transients in Section \ref{sec:sed_fitting}, and our galaxy SED fitting technique for the Warhol arc in Section \ref{sec:prospector}. Our stellar population synthesis and microlensing simulations are presented in Section \ref{sec:simulation}, and we discuss our results and compare the properties of the observed stars with the simulation in Section \ref{sec:results}. Section \ref{sec:conclusion} summarizes our conclusions.

Throughout this paper, we assume a flat $\Lambda$-cold-dark-matter cosmology with matter density parameter $\Omega_{M}=0.287$ and Hubble parameter H$_0=69.3$~km~s$^{-1}$~Mpc$^{-1}$. These cosmological parameters were derived from the nine-year Wilkinson microwave anisotropy probe \citep[WMAP9;][]{Hinshaw_2013}. All dates and times are reported in UTC.

\section{Observations and Data Reduction}\label{sec:observations}
\subsection{JWST Observations}\label{subsec:nircam}
The M0416 cluster field was observed in four separate visits separated by a total of 126 days with NIRCam on {\it JWST}. Three visits were taken by the PEARLS program and one by the CANUCS program. All four visits obtained imaging in the same eight bands: F090W, F115W, F150W, F200W, F277W, F356W, F410M, and F444W. The exposure time for each PEARLS visit was 48.7--62.9\,min per filter, and the exposure time for the CANUCS visit was 106.6\,min in each filter. The dates, position angles, and exposure times for each visit are given in Table~\ref{tab:obs}.

\begin{table}
    \caption{NIRCam Exposure Times and Position Angles}
    \centering
    \ra{1.2}
    \begin{tabular}{l|l}
    \toprule \toprule
        {\bf Visit 1 -- 2022 Oct 7}  & {\bf Visit 2 -- 2022 Dec 29}\\
        PEARLS; PA = 293$^{\rm o}$ & PEARLS; PA = 33$^{\rm o}$ \\
        F090W/F444W: 3779.343~s  &F090W/F444W: 3779.343~s\\
        F115W/F410M: 3779.343~s & F115W/F410M: 3779.343~s\\
        F150W/F356W: 2920.401~s & F150W/F356W: 2920.401~s\\
        F200W/F277W: 2920.401~s & F200W/F277W: 2920.401~s  \\

        \midrule
        {\bf Visit 3 -- 2023 Jan 11} & {\bf Visit 4 -- 2023 Feb 10}\\
        CANUCS; PA = 49$^{\rm o}$ & PEARLS; PA = 71$^{\rm o}$ \\
        F090W/F444W: 6399.115~s & F090W/F444W: 3779.343~s \\
        F115W/F410M: 6399.115~s& F115W/F410M: 3349.872~s \\
        F150W/F356W: 6399.115~s & F150W/F356W: 2920.401~s \\
        F200W/F277W: 6399.115~s& F200W/F277W: 2920.401~s \\
    \bottomrule
    \end{tabular}
    
    \label{tab:obs}
\end{table}

\subsection{JWST Data Reduction}
The PEARLS and CANUCS NIRCam observations were downloaded from the Mikulski Archive for Space Telescopes (MAST). We retrieved the Stage 1 data products from MAST, which are single NIRCam exposures that have been processed by the Stage 0 preprocessing pipeline. We reduce these data using version 1.15.0 of the public {\it JWST} science calibration pipeline \citep{bushouse_2023} with reference files from {\tt JWST\_1253.pmap}. Stage 1 (detector-level corrections) and Stage 2 (photometric calibration) of the pipeline are both run with all default parameters. Stage 3 of the pipeline coadds the individual calibrated exposures to produce science-level mosaics. Using the Stage 3 pipeline, we resample the mosaics to a 0.03$\arcsec$ pixel scale and project all epochs onto a common pixel grid. Color-composite images for all four epochs are shown in Figure \ref{fig:warhol_cutouts}. The {\it JWST} data used in this paper can be found in MAST: \dataset[10.17909/7rqz-qy32]{http://dx.doi.org/10.17909/7rqz-qy32}.

\subsection{HST Flashlights Observations}
M0416 has been targeted by {\it HST} as part of the Flashlights program \citep{Kelly_2022}. Flashlights obtained two epochs of deep imaging of M0416 in the F200LP and F350LP filters. Epoch 1 was obtained on 2020 October 1 and epoch 2 2021 on October 12, a baseline of $\sim 1$\,yr. The 5$\sigma$ depth was 30.0\,mag for both epochs and both filters. Four transients were detected in the difference images between the two Flashlights epochs (Kelly et al. 2025, in prep.). We refer to those four Flashlights transients as F1, F2, F3, and F4 in the remainder of this work. The coordinates of the Flashlights transients are listed in Table \ref{tab:transients}.

\subsection{Keck MOSFIRE Spectroscopy}\label{subsec:MOSFIRE}

\begin{figure*}
\centering
    \includegraphics[width=\linewidth]{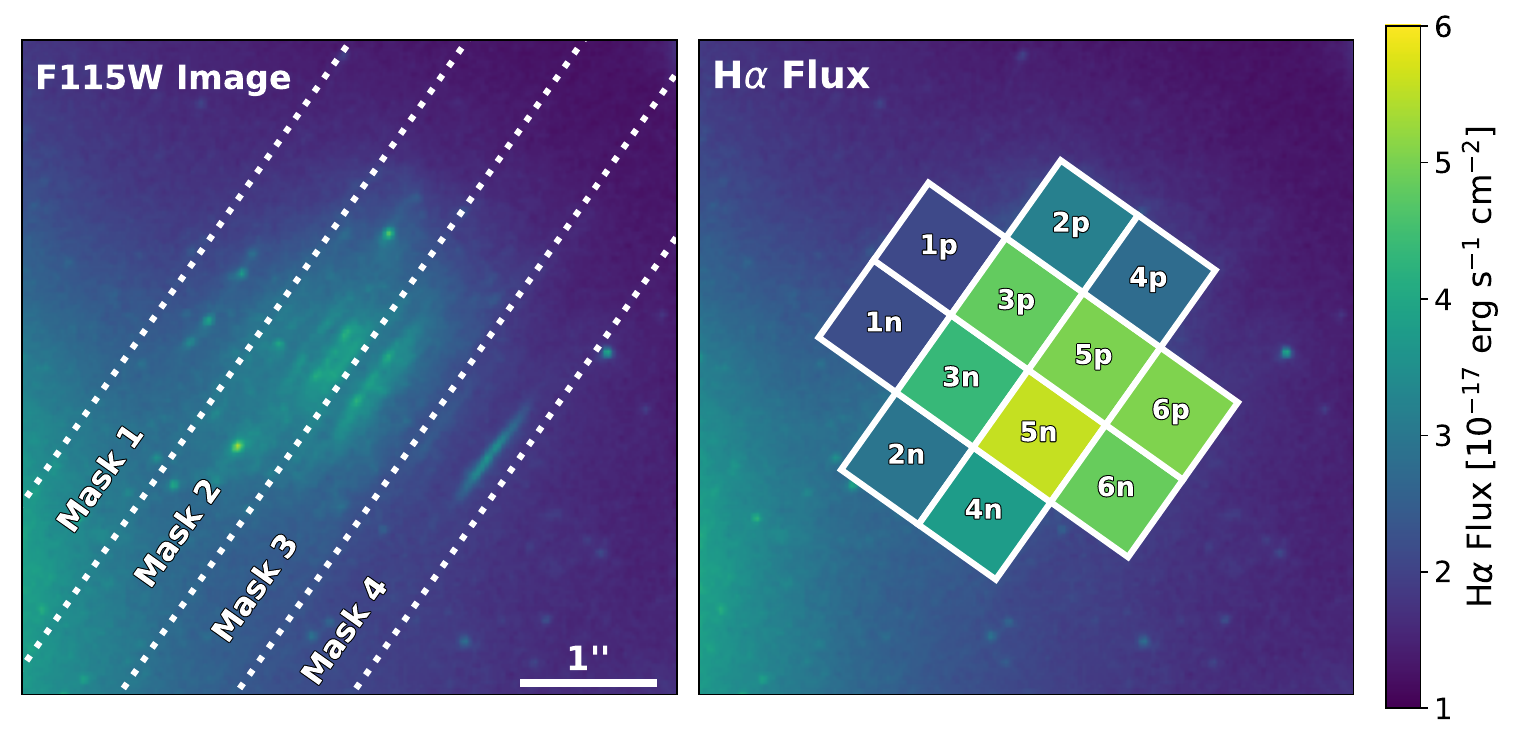}
    \caption{{\it Left:} Slits used for our Keck MOSFIRE observations of the Warhol arc, shown on the {\it JWST} NIRCam F115W image of the galaxy. Each slit is 0.7$\arcsec$ wide. {\it Right:} H$\alpha$ flux measured from the MOSFIRE spectroscopy in twelve 0.7$\arcsec \times 0.7\arcsec$ boxes. Boxes labeled with ``n'' are on the negative-parity side of the critical curve, and boxes labeled with ``p'' are in the symmetrical position on the positive-parity side of the critical curve. {The orientation of these images is North pointing upward and East pointing to the left.}}
    \label{fig:mosfire_slits}
\end{figure*}

On 2022 November 15, we acquired 104\,min of near-IR (NIR) spectroscopy of the Warhol arc using the Multi-Object Spectrometer for Infra-Red Exploration (MOSFIRE; \citealt{McLean}) on the Keck-I 10\,m telescope (PI A. V. Filippenko). We used the \textit{J} filter which provides a spectral resolving power of $R \approx 3300$ and has the H$\alpha$ emission line within its spectral coverage (1.153--1.352\,$\mu$m) at the redshift of the Warhol arc ($z=0.94$). A 0.7$\arcsec$-wide slit was aligned perpendicular to the critical curve and centered the slit on the Warhol arc. We conducted a series of four observations, shifting the slit along the arc by its width (0.7$\arcsec$) each time to obtain spectra of the entire arc (see Figure \ref{fig:mosfire_slits}). 

{An ABBA dither pattern with a 24$\arcsec$ nod amplitude was used with 120\,s exposures.} The total exposure time was 24\,min each for Mask 1, Mask 3, and Mask 4, and 32\,min for Mask 2. We included two field stars on each mask to use for flux calibration. Following our observations of the Warhol arc, we acquired a spectrum of the telluric A0~V standard star HIP~13917 using a 0.7$\arcsec$-wide slit. {The average seeing full width at half-maximum intensity (FWHM) was 1.0$\arcsec$.}

The data were reduced using the MOSFIRE data-reduction pipeline \citep[DRP;][]{Konidaris_2019}. The DRP used dome-flat exposures to apply a flat-field correction and trace the slit edges. Bright OH night-sky lines from the science exposures were used to generate the wavelength solution and rectify the slits. The pipeline created sky-subtracted spectra by subtracting the ``B" frames from the ``A" frames (and vice versa), and coadded all frames to produce the two-dimensional (2D) science spectrum for each slit.  

We used the telluric star HIP~13917 to correct for the detector's relative response along the wavelength axis; its observed  spectrum was compared with a synthetic Vega spectrum generated using {\tt synphot} \citep{synphot} to estimate the relative response function. For the absolute flux calibration {and to correct for slit losses}, we selected a field star on each mask and compared its $J$-band flux to {\it HST} photometry of the star in the WFC3-IR F125W filter. We multiplied the science spectra by the relative response function and the absolute flux-calibration factor to obtain flux-calibrated spectra of the Warhol arc.

\subsection{LBT LUCI Spectroscopy}\label{subsec:LUCI}
On 2023 December 11, we obtained NIR spectra of the Warhol arc using the twin LBT Utility Camera in the Inrafred (LUCI; \citealt{LUCI_instrument}) instruments on the Large Binocular Telescope (LBT). We used a 1.5$\arcsec$-wide long slit centered on the arc and the G210 grating in the \textit{z} and \textit{J} filters. The spectral range of the \textit{z} filter (0.762--1.152\,$\mu$m) contains the H$\beta$, [\ion{O}{3}]$\lambda$4959, and [\ion{O}{3}]$\lambda$5007 emission lines, and the \textit{J} filter (0.942--1.552\,$\mu$m) contains the [\ion{N}{2}]$\lambda$6548, H$\alpha$, and [\ion{N}{2}]$\lambda$6583 emission lines. The G210 grating provides a spectral resolving power $R\approx5400$ in the $z$ band and $R\approx5800$ in the \textit{J} band.  {We used an ABBA dither pattern with a 30$\arcsec$ nod amplitude and 180\,s exposures.} The total integration time was 96\,min in each filter. The full width at half-maximum intensity (FWHM) of the seeing was $\sim 0.7\arcsec$. We acquired spectra of the telluric standard star HIP~13917 using the same long-slit mask immediately following our observations of Warhol.  

The data were reduced using {\tt PypeIt} \citep{pypeit_citation}. The pipeline performs dark and bias subtraction, and uses dome-flat exposures to measure and apply a flat-field correction. Cosmic rays are identified and masked using Laplacian edge detection \citep{vanDokkum_2001}. Wavelength calibration is performed using bright OH night-sky lines in the science frames. The pipeline subtracts the ``B" dither position frames from the ``A" dither position frames (and vice versa) to create sky-subtracted science spectra. We extract the 1D spectra of the Warhol arc and the standard star using a 1.5$\arcsec$-wide boxcar aperture. 

We use the observed standard-star spectrum to perform flux calibration on the science data. HIP~13917 is an A0~V star, so we generate a synthetic Vega spectrum scaled to the star's magnitudes in the $z$ and $J$ bands using {\tt synphot} \citep[][]{synphot}. We calculate the wavelength-dependent flux-calibration function by dividing the synthetic Vega spectrum by the observed stellar spectrum. The science spectra are then multiplied by this flux-calibration function.

To correct for the emission that falls outside of the 1.5$\arcsec \times 1.5\arcsec$ aperture, we multiply the LUCI spectrum by the ratio of the total H$\alpha$ flux measured from the Keck MOSFIRE spectra to the H$\alpha$ flux measured in the smaller LUCI aperture. See Section \ref{subsec:linefluxes} for details on emission-line flux measurements.

\subsection{MUSE Spectroscopy}
We use archival integral field unit (IFU) optical spectroscopy of the Warhol arc from the Multi Unit Spectroscopic Explorer (MUSE) spectrograph on the Very Large Telescope (VLT). Deep MUSE IFU spectroscopy of M0416 was acquired over 16 observations from 2017 November through 2019 August (Program ID 0100.A-0763(A); PI E. Vanzella). The total integration time was 15.8\,hr and the final stacked seeing FWHM was 0.74$\arcsec$. The spectral range of MUSE is 0.46--0.93\,$\mu$m, and the spectral resolving power ranges from $R\approx2000$ at 0.46\,$\mu$m to $R\approx4000$ at 0.93\,$\mu$m \citep{Bacon_2010}.

\begin{figure}
    \centering
    \includegraphics{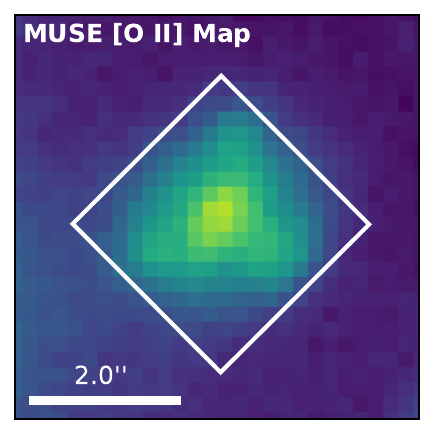}
    \caption{[MUSE \ion{O}{2}] $\lambda$3727 flux-intensity map of the Warhol arc. The white box shows the 2.8$\arcsec$ square aperture used to extract the 1D spectrum. {The orientation of this image is North pointing upward and East pointing to the left.}}
    \label{fig:MUSE}
\end{figure}

We obtained the fully reduced MUSE datacube of M0416 from the ESO Science Archive \footnote{http://archive.eso.org/scienceportal/home}. The 1D optical spectrum of the Warhol arc was extracted using a 2.8$\arcsec$ square aperture. We extracted the sky spectrum in five nearby positions using the same 2.8$\arcsec$ aperture and created a sky spectrum from the median of those five sky extractions, which we subtracted from the Warhol science spectrum to create the final, sky-subtracted 1D MUSE spectrum of the Warhol arc. At $z=0.94$, the MUSE wavelength coverage in the rest frame is 0.89--1.80\,$\mu$m. Figure \ref{fig:MUSE} shows an [\ion{O}{2}] $\lambda$3727 flux-intensity map of the Warhol arc with the 2.8$\arcsec$ square extraction aperture displayed on top.

\section{Transient Search}\label{sec:transient_search}

We conduct a search for transients in the Warhol galaxy by detecting peaks in the difference images between epochs. The search is performed on the difference images created by direct subtraction from the following pairs of visits: V1--V2, V1--V3, V1--V4, V2--V3, and V2--V4. The initial search is done using the F277W difference images. {We repeat the search on all eight filters to ensure that we do not miss hot 
sources that are bright at short wavelengths but undetected at long wavelengths, but the other filters do not reveal any additional transients beyond those found in the F277W search.} 

We begin by identifying all positive and negative peaks in a 3.0$\arcsec \times 3.0\arcsec$ box centered on the Warhol arc. Next, we measure the flux of each candidate using our point-spread-function (PSF) fitting photometry method (see Section \ref{subsec:photometry} for a full description of this method). We reject all candidates with signal-to-noise ratio (S/N) $<$ 5 and proceed with all candidates with S/N $\geq 5$. Next, we repeat our PSF-fitting photometry on the candidates in the other three long-wavelength filters (F356W, F410M, F444W). We require that each candidate has S/N $\geq 5$ in at least one other filter. Lastly, we visually inspect the remaining candidates to form our final list of robust transients. 

Using this detection method, we recover all seven transients in the Warhol region that were reported by Y23. We denote these sources as W1--W7, matching the nomenclature of Y23. Additionally, we find two new transients in the Warhol arc that were not found in the original transient search by Y23; we denote these two as W8 and W9. The nine Warhol arc transients are shown in Figure \ref{fig:warhol_cutouts}, and their positions are listed in Table \ref{tab:transients}.

\begin{table}
    \caption{Warhol Transient Positions}
    \label{tab:transients}
    \centering
    \ra{1.25}
    \begin{tabular}{clll}
    \toprule \toprule
    \bf Name & \bf ra (deg) & \bf dec (deg) & \bf Visits Detected\\
    \midrule
    W1 & 64.0370208 & -24.06727083 & V1,V2,V3,V4\\
    W2 & 64.0368223 & -24.06726380 & V1,V2,V3,V4\\
    W3 & 64.0367250 & -24.06730472 & V2,V3 \\
    W4 & 64.0366250 & -24.06732722 & V3\\
    W5 & 64.0367653 & -24.06735219 & V4\\
    W6 & 64.0366214 & -24.06747723 & V4\\
    W7 & 64.0365712 & -24.06738799 & V4\\
    W8 & 64.0365282 & -24.06737374 & V2, V3\\
    W9 & 64.0366583 & -24.06725306 & V2\\
    \midrule
    F1$^a$ & 64.0363046 & -24.0675050 & Flashlights \\
    F2$^a$ & 64.0365524 & -24.0673339 & Flashlights\\
    F3$^a$ & 64.0364386 & -24.0674422 & Flashlights \\
    F4$^a$ & 64.0365133 & -24.0673656 & Flashlights\\
    \bottomrule
    
    \end{tabular}
    \footnotesize{{\bf Notes.} Transients W1$-$W7 were reported by \cite{Yan_2023}.\\}
    \footnotesize{$^a$ Coordinates from Kelly et al. (2025, in prep.)}
\end{table}

We compute the 5$\sigma$ detection limit in each filter by injecting simulated sources into the V4--V1 difference image in positions near the Warhol arc. The simulated sources are point sources convolved with the PSF of each filter. We repeatedly increase the flux of the simulated source until it can be detected at the 5$\sigma$ level with the same PSF-fitting photometry method described in Section \ref{subsec:photometry}. The 5$\sigma$ detection limits are given in Table \ref{tab:detection}.

\begin{table}
    \caption{5$\sigma$ Detection Limits}
    \label{tab:detection}
    \centering
    \ra{1.25}
    \setlength{\tabcolsep}{15pt}
    \begin{tabular}{ll|ll}
    \toprule \toprule
    \bf SW& \bf Limit & \bf LW & \bf Limit\\
    \bf Filter & \bf  (mag)& \bf Filter & \bf  (mag)  \\ 
    \midrule
          F090W & 29.7 & F277W& 29.5\\
          F115W & 29.5  &F356W& 29.6 \\ 
          F150W & 29.5 &F410M &29.0 \\ 
          F200W & 29.5 & F444W & 29.3 \\
         \midrule
    \end{tabular}
    \footnotesize{{\bf Notes. } Magnitudes are given in the AB system \citep{OkeGunn_1983}.}
\end{table}
\section{Measurements}\label{sec:measurements}
\subsection{Photometry}\label{subsec:photometry}

We perform PSF-fitting photometry on the {\it JWST} NIRCam images to measure the fluxes of the nine identified transients in the Warhol arc. Using {\tt photutils} \citep{photutils}, we construct an effective PSF \citep[ePSF;][]{Anderson_2000} for each NIRCam filter using eight unsaturated, high-S/N, isolated stars in the M0416 field. The ePSF is constructed using a least-squares fitting routine which creates a model for the fraction of the flux of a point source that lands within each pixel.

We use the ePSFs to measure the fluxes of the Warhol transients using PSF-fitting photometry. We fit the ePSF model to each transient using a nonlinear least-squares routine to infer the flux and centroid of the source. We fit each source within a square box whose length is approximately twice the FWHM of the NIRCam filter. In cases of low S/N, we first measure the centroid of the source in a higher S/N filter and then fix the source's centroid to that position for the low S/N filters. 

The background is estimated by calculating the median flux within a circular annulus with an area of 0.4 square arcsec centered on the source, with the source itself and any other transients within the field of view masked out. Since the transients are embedded in the Warhol galaxy, the background is highly nonuniform and the uncertainty is background dominated. To account for this systematic uncertainty, we inject a synthetic source with a flux equal to the measured flux of the transient into 150 nearby positions. We then recover the fluxes of the injected sources using the PSF-fitting photometry method, and we use the standard deviation of the recovered fluxes as the systemic uncertainty associated with the nonuniform background.


The PSF-fitting photometry is performed on the difference images between the epoch(s) in which the source is detected and the epoch(s) in which the source is not detected. Transients W1 and W2 are detected in all four epochs, so we measure the photometry without difference imaging in the epoch in which the transient is the faintest (Visit 1 for source W2 and Visit 4 for source W1), and then measure the photometry in the difference images between the other three epochs and the faintest epoch. Flux measurements for each transient are shown in Table \ref{tab:photometry}




%

\subsection{Emission-Line Flux Measurements}\label{subsec:linefluxes}

\begin{table}
    \caption{Warhol Emission-Line Fluxes}
    \label{tab:line_fluxes}
    \centering
    \ra{1.3}
    \setlength{\tabcolsep}{7pt}
    \begin{tabular}{cccc}
    \toprule\toprule
         \multicolumn{4}{c}{\textbf{VLT MUSE}}\\
         \textbf{Line}&$\mathbf{\lambda_{rest}}$ &\textbf{$\mathbf{F_{obs}}^a$}&\textbf{$\mathbf{F_{corr}}^a$} \\
         \midrule

        [\ion{O}{2}] & 3727 & 26.8 $\pm$ 0.24 & 38.2 $\pm$ 2.5\\
        
        [\ion{Ne}{3}] & 3869 & 1.28 $\pm$ 0.06 & 1.86 $\pm$ 0.15\\
        
        H$\delta$ & 4102 & 1.78 $\pm$ 0.21 & 2.62 $\pm$ 0.25\\
        
        H$\gamma$ & 4340 & 3.26 $\pm$ 0.38 & 4.72 $\pm$ 0.45\\

        \midrule

        \multicolumn{4}{c}{\textbf{LBT LUCI}}\\
        
         \textbf{Line}&$\mathbf{\lambda_{rest}}$ &\textbf{$\mathbf{F_{obs}}^a$}&\textbf{$\mathbf{F_{corr}}^a$} \\

         \midrule
         
        [\ion{O}{3}] & 5007 & 8.06 $\pm$ 2.16 & 11.0 $\pm$ 3.0\\
        H$\alpha$ & 6563 & 22.9 $\pm$ 1.9 & 28.7 $\pm$ 2.7\\
        
        \midrule

        \multicolumn{4}{c}{\textbf{Keck MOSFIRE}}\\
    \textbf{Box} & \textbf{H$\alpha$ Flux}$^a$ & \textbf{H$\alpha$ Flux}$^a$ & \textbf{Flux Ratio}\\
    \textbf{IDs}& \textbf{\scriptsize{(neg. parity)}} & \textbf{\scriptsize{(pos. parity)}} &\textbf{\scriptsize{(neg./pos.)}}\\
    \midrule
         1n, 1p & 1.08 $\pm$ 0.36 & 1.04 $\pm$ 0.30 &1.0 $\pm$ 0.5  \\
         2n, 2p & 1.52 $\pm$ 0.23 & 1.63 $\pm$ 0.24 & 0.9 $\pm$ 0.2  \\
         3n, 3p & 2.24 $\pm$ 0.35 & 2.47 $\pm$ 0.26 & 0.9 $\pm$ 0.2 \\
         4n, 4p & 1.93 $\pm$ 0.35 & 1.41 $\pm$ 0.32 & 1.4 $\pm$ 0.4\\
         5n, 5p & 2.86 $\pm$ 0.39 & 2.58 $\pm$ 0.49 & 1.1 $\pm$ 0.3\\
         6n, 6p & 2.40 $\pm$ 0.45 & 2.49 $\pm$ 0.47 & 1.0 $\pm$ 0.3\\
        
         \bottomrule
    \end{tabular}
    \footnotesize{$a$: Fluxes reported in units of $10^{-17}$\,erg\,s$^{-1}$\,cm$^{-2}$}
\end{table}


\begin{figure*}[ht!]
   \centering
   \includegraphics[width=\linewidth]{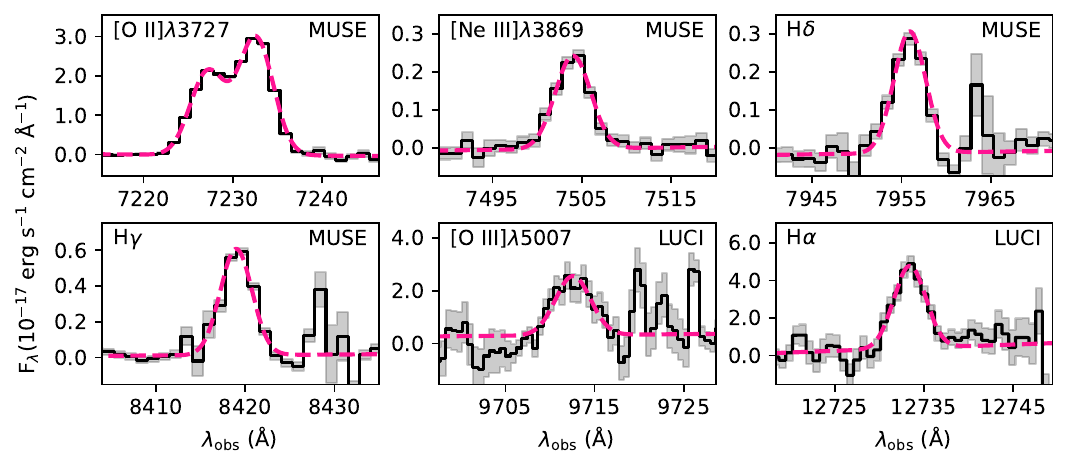}
   \caption{MUSE spectroscopy of the [\ion{O}{2}] $\lambda$3727, [\ion{Ne}{3}] $\lambda$3869, H$\delta$, and H$\gamma$ emission lines, along with LUCI spectroscopy of the [\ion{O}{3}] $\lambda$5007, and H$\alpha$ emission lines, in the Warhol arc. The black line shows the spectrum in a 30\,\r{A} window centered on each emission line, and the shaded gray region shows the 1$\sigma$ uncertainty interval. The dashed pink lines show our best-fitting Gaussian profiles.}
   \label{fig:emission_lines}
\end{figure*}

    

We measure the emission-line fluxes of [\ion{O}{2}] $\lambda$3727, [\ion{Ne}{3}] $\lambda$3869, H$\delta$, and H$\gamma$ from the MUSE spectroscopy, and the fluxes of [\ion{O}{3}]$ \lambda$5007 and H$\alpha$ from the LUCI spectroscopy.  We use a nonlinear least-squares function to fit a Gaussian profile to each of the emission lines within a 30~\r{A} window centered on the central wavelength of the emission line in the observer frame. The [\ion{O}{2}]$ \lambda$3727 doublet is modeled simultaneously as a blended profile of two superimposed Gaussians. 

We model the three detected Balmer lines (H$\delta$, H$\gamma$, and H$\alpha$) simultaneously by fitting for the flux of H$\alpha$ and the Warhol arc's internal attenuation due to dust in the $V$ band ($A_V$). The dust-corrected fluxes of H$\gamma$ and H$\delta$ ($F_{{rm H}\gamma,{\rm corr}}$ and $F_{{\rm H}\delta,{\rm corr}}$) are set by the theoretical Case B recombination ratios from \cite{Osterbrock},

\begin{equation}
    \frac{F_{{\rm H}\gamma , {\rm corr}}}{F_{{\rm H}\alpha , {\rm corr}} } = 0.164\, ,  \hspace{10mm}  \frac{F_{{\rm H}\delta , {\rm corr}}}{F_{{\rm H}\alpha , {\rm corr}}} = 0.091\, .
\end{equation}

The best-fitting dust attenuation inferred from the H$\delta$, H$\gamma$, and H$\alpha$ fluxes is $A_V = 0.30 \pm 0.18$\,mag. The dust-corrected fluxes are converted to the observed (reddened) fluxes $F_{\rm obs}$ using the inferred value of $A_V$ and assuming a \cite{Cardelli_1989} attenuation curve with $R_V=3.10$. Table \ref{tab:line_fluxes} lists the observed and dust-corrected flux of each detected emission line and Figure \ref{fig:emission_lines} shows the LUCI spectroscopy and best-fitting Gaussian profiles. {At $z=0.94$, H$\beta$ falls near the edge of the LUCI detector where sensitivity is relatively low, and we do not detect H$\beta$ in the $z$-band LUCI spectrum. The 3$\sigma$ upper limit on the H$\beta$ flux is $15.3\times10^{-17}$ erg s$^{-1}$ cm$^{-2}$.}

We use the dust-corrected emission-line fluxes to infer the nebular oxygen abundance of the Warhol arc. We apply the popular ``R23" empirical strong-line metallicity calibration, where

\begin{equation}
    R23 \equiv \frac{F([{\rm O II}] \lambda3727) + F([{\rm O III}] \lambda\lambda4959,5007)} {F(H\beta)}\, .
\end{equation}

The dust-corrected flux of [\ion{O}{3}] $\lambda$4959 is computed according to its flux ratio compared to [\ion{O}{3}] $\lambda$5007, set to 0.33 by atomic physics \citep{Storey_2000}. The dust-corrected flux of H$\beta$ is computed from the dust-corrected H$\alpha$ flux assuming Case B recombination. {Using a calibration for the R23 metallicity based on a sample of low-redshift star-forming galaxies \citep{Jones_2015}, we infer an oxygen abundance for the Warhol arc, $12+
\log({\rm O}/{\rm H}) = 8.45\pm0.08$.}

{We test how the inferred oxygen abundance changes with different strong-line metallicity indicators. Using the calibrations from \cite{Jones_2015} and the dust-corrected emission-line fluxes, the oxygen abundance is computed from the  indicators [\ion{Ne}{3}]/[\ion{O}{2}], [\ion{O}{3}]/[\ion{O}{2}], and [\ion{O}{3}]/H$\beta$. All three indicators yield oxygen abundances that are consistent with the value derived from the R23 indicator within the combined 1$\sigma$ uncertainties ($12+\log({\rm O}/{\rm H})=8.43\pm0.22$, $8.59\pm0.24$, and $8.58\pm0.15$, respectively). We also test a different calibration for the R23 indicator, derived by \cite{Curti_2020} from a sample of low-redshift star-forming galaxies. The oxygen abundance inferred by this calibration is $12+\log({\rm O}/{\rm H})=8.55\pm0.13$, consistent with the result from the \cite{Jones_2015} R23 calibration.}

The dust-corrected flux ratio of [\ion{O}{3}]/[\ion{O}{2}] can be used to infer the ionization parameter of the gas in the Warhol arc. Using a calibration based on photoionization models from \cite{Levesque_2014}, we infer $\log(U)=-3.30\pm0.12$ for the Warhol galaxy.

\subsubsection{H$\alpha$ Emission Symmetry}
The multislit Keck MOSFIRE spectroscopy is used to test for magnification asymmetries on opposite sides of the critical curve. We measure the H$\alpha$ flux in 12 $0.7\arcsec \times 0.7\arcsec$ boxes positioned symmetrically on either side of the critical curve (see Figure~\ref{fig:mosfire_slits}). The H$\alpha$ fluxes of all pairs of matching boxes on each side of the critical curve are consistent with each other within the 1$\sigma$ uncertainties (see Table~\ref{tab:line_fluxes}). Therefore, the measurements do not indicate any magnification asymmetry for these apertures across the critical curve for the Warhol arc.

\section{Transient SED Fitting}\label{sec:sed_fitting}

\begin{figure*}
    \centering
    \includegraphics[width=.9\linewidth]{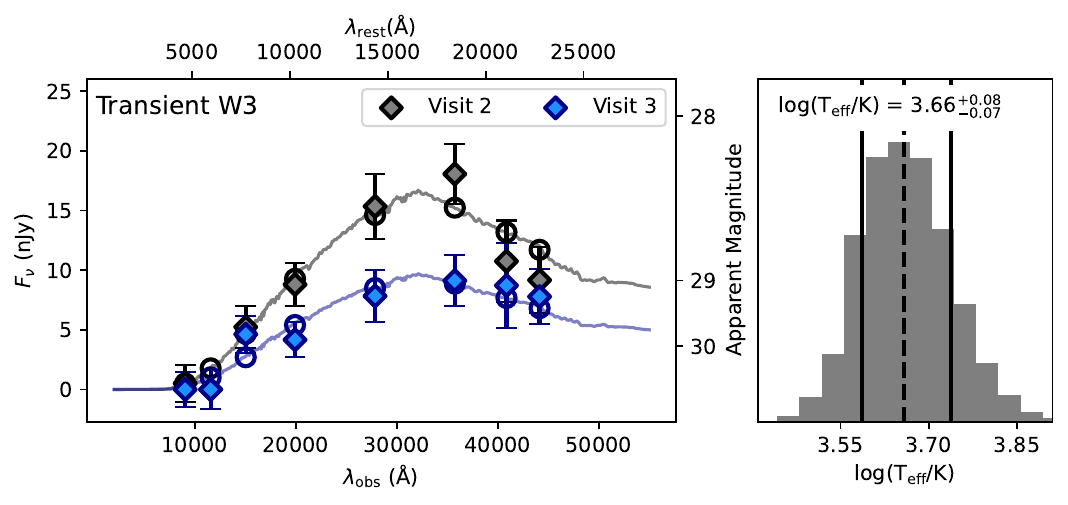}
    \caption{{\it Left:} Observed photometry and best-fitting stellar model for transient source W3. The shaded diamonds show the observed photometry and the open circles indicate the synthetic photometry from the best-fitting model. {\it Right:} The posterior of the logarithm of the stellar temperature for source W3.}
    \label{fig:W3_model}
\end{figure*}

We use Bayesian inference to find a best-fitting stellar model for the SED of each transient. The free parameters are the star's flux in the F277W filter $F_{\rm F277W}$, effective temperature $T_{\rm eff}$, surface gravity $\log(g)$, stellar metallicity $Z_*/Z_{\odot}$, line-of-sight extinction due to dust $A_V$, and  ratio of total-to-selective extinction $R_V$.  
We use {\tt pystellibs}\footnote{https://github.com/mfouesneau/pystellibs} with the BaSeL stellar library \citep{Lejeune_1998} to interpolate stellar atmosphere models and generate model stellar spectra for a given temperature, surface gravity, and stellar metallicity. Using a dust curve function for arbitrary $R_V$ \citep{Fitzpatrick_1999}, we apply the wavelength-dependent extinction correction to the generated model spectrum for a given $R_V$ and $A_V$. We use {\tt pysynphot}\footnote{https://github.com/spacetelescope/pysynphot} to rescale the spectrum based on the $F_{\rm f356w}$ model parameter and generate synthetic NIRCam photometry in each filter from the rescaled spectrum.

The probability density for a given set of model parameters $\theta$ is computed using the multivariate Gaussian probability density,
\begin{equation}
    p(\theta) \propto {\rm exp} \left [-\frac{1}{2}(\vec{x}(\theta) - \vec{\mu})^{\rm T} \Sigma^{-1} (\vec{x}(\theta) - \vec{\mu}) \right ]\, ,
\label{eqn:probability}
\end{equation}
where $\vec{x}(\theta)$ is the vector of synthetic NIRCam photometry in each filter, $\vec{\mu}$ is the vector of the observed NIRCam photometry, and $\Sigma$ is the covariance matrix. We estimate the off-diagonal terms of the covariance matrix using the 150 recovered flux residuals from our PSF-fitting photometry procedure (see Section \ref{subsec:photometry}).

The MCMC ensemble sampler {\tt emcee} \citep{emcee} is used to draw samples from the multivariate Gaussian probability density and explore the parameter space. A uniform prior is applied for each free parameter (see Table \ref{tab:starfit}). We burn in the sample using 32 walkers and 100 steps. After the burn-in, we allow the {\tt emcee} sampler to explore the parameter space using 32 walkers and 10,000 steps. We use the median of the resulting posteriors as the best-fit parameter values, and the 84\% confidence interval as the 1$\sigma$ uncertainties.

\begin{table}
    \caption{Stellar SED-Fitting Parameters}
    \label{tab:starfit}
    \centering
    \ra{1.3}
    \begin{tabular}{cll}
    \toprule\toprule

    \textbf{Parameter} & \textbf{Prior} & \textbf{Units}\\
    \midrule
         $F_{\rm F277W}$& Uniform [0, 1000] & nJy  \\
         $\log(T_{\rm eff})$& Uniform [3.2, 4.8] & $\log({\rm K})$ \\
         $\log(g)$ & Uniform [-2, 5] & $\log({\rm cm\,s}^{-1})$\\
         $Z_*/Z_{\odot}$ & Uniform [0.01, 2.0] & dimensionless\\
         $A_{\rm V}$ & Uniform [0, 4] & AB magnitude \\
         $R_{\rm V}$ & Uniform [0, 10] & dimensionless \\
     \bottomrule
    \end{tabular}

\end{table}

For the transients that are visible at more than one epoch, we model the photometry in all epochs simultaneously, allowing the star's apparent magnitude to vary between each epoch as the magnification changes. The values of $T_{\rm eff}$, $\log(g)$, $Z_*/Z_{\odot}$, $A_V$, and $R_V$ are consistent between all epochs, since we assume we are observing the same star with a different magnification factor. {For the transients that are visible in all four epochs, the model spectrum at the faintest epoch is subtracted from the model spectra at the other epochs to accurately match the difference photometry for these sources (see Section \ref{subsec:photometry}).}

\begin{figure}
    \centering
    \includegraphics[width=\linewidth]{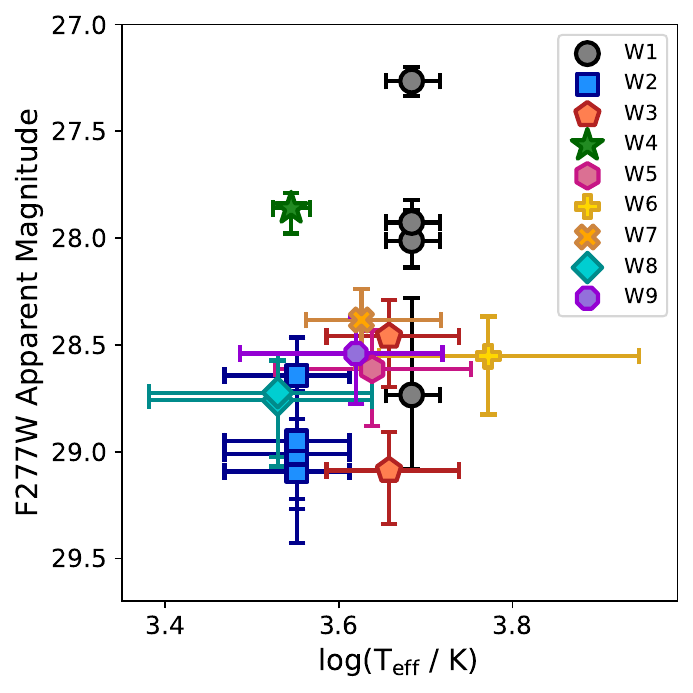}
    \caption{Best-fitting effective temperatures and observed F277W apparent magnitudes for the nine Warhol transients. The transients that appear more than once in this plot are detected at multiple epochs. The apparent magnitudes are not corrected for magnification.}
    \label{fig:magT}
\end{figure}

Figure \ref{fig:W3_model} shows an example of the best-fitting stellar model spectrum for source W3. Model spectra for the other eight transient sources are shown in the Appendix in Figure \ref{fig:W1W2W4W5W6W7W8W9_model}. A color-magnitude diagram indicating the best-fit temperatures and observed F277W apparent magnitudes at each epoch is shown in Figure \ref{fig:magT}, and the best-fitting parameter values are listed in \ref{tab:best_fits}.

We also fit each source with a binary system of two stars with different temperatures and different surface gravities. The fitting method is the same as described above for the single-star case, with three extra free parameters for the companion star -- its temperature, surface gravity, and F277W flux. The two model stellar spectra are then added together before generating the synthetic NIRCam photometry in each filter.

To test whether the binary model produces a better fit accounting for the increased number of model parameters, we use the Bayesian Information Criterion (BIC). When comparing the goodness-of-fit of two models, the model with the lower BIC is preferred. The BIC is given by
\begin{equation}
   {\rm BIC} = k \ln(N) - 2\ln(\hat{L})\, ,
\end{equation}
\noindent where $k$ is the number of model parameters, $N$ is the number of observed data points, and $\hat{L}$ is the maximum value of the likelihood function \citep{schwarz1978estimating}. 

The difference in the BIC 
between the single-star model and the binary model is computed for each transient,
\begin{equation}
    \Delta {\rm BIC} \equiv {\rm BIC}_{\rm single} - {\rm BIC}_{\rm binary}\, ,
\end{equation}

\noindent and the binary model produces an improved fit compared to the single-star model for two of the transients, W1 and W2. For source W1, $\Delta {\rm BIC}  = 6.5$, and for source W2, $\Delta {\rm BIC}  = 26.7$. While the $\Delta {\rm BIC}$ value for source W1 only suggests ``moderate" evidence in favor of the binary model, the result of $\Delta {\rm BIC}>10$ for source W2 suggests ``very strong" evidence that the binary model is preferred. {The other seven transients have negative $\Delta {\rm BIC}$ values and therefore show no evidence for a companion.}

The light curve of source W2 exhibits a dramatic change in color over the 126\,day light curve, where the blue component grows brighter over time while the red component becomes fainter. This change in color suggests that the binary system may be orbiting across the critical curve of the microlensing caustic. \cite{Williams_2025_BINARY} presents an analysis of the possible binary-star parameters and orbital configuration that could reproduce this light curve.

\section{Galaxy SED Fitting}\label{sec:prospector}
\begin{figure*}
    \centering
    \includegraphics[width=\linewidth]{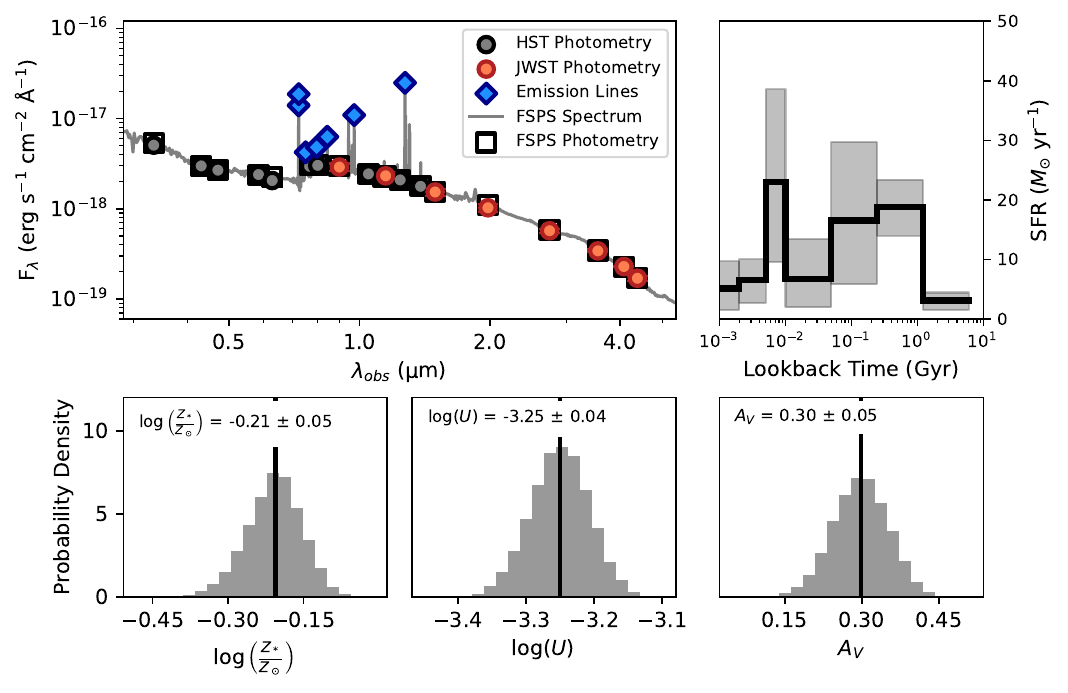}
    \caption{{\it Left:} Best-fitting model SED for the Warhol arc from our {\tt fsps} fit. The gray circles show the {\it HST} photometry of the arc, the orange circles indicate the {\it JWST} photometry, and the blue diamonds represent the observed emission lines. The gray line shows the best-fitting composite spectrum from {\tt fsps}, and the open squares give the best-fitting model photometry. {\it Right:} Best-fitting SFH for the Warhol arc from the {\tt fsps} fitting. The black line shows the best-fit SFR in each temporal bin and the gray shading provides the 1$\sigma$ uncertainty. These measurements are not corrected for magnification.}
    \label{fig:bestfit_235}
\end{figure*}

We use the Flexible Stellar Population Synthesis package \citep[{\tt fsps};][]{Conroy_2009,Conroy_2010} to generate simulated galaxy spectra and model the SED of the Warhol arc. The photometry of the arc is measured from BCG-subtracted and PSF-matched {\it HST} images from \cite{Shipley_2018} in 13 {\it HST} filters spanning a wavelength range of 0.2\,$\mu$ to 1.4\,$\mu$m. We create BCG-subtracted {\it JWST} NIRCam images using {\tt galfit} \citep{Peng_2010} to model the disk profile of the BCG in each filter and measure the photometry of the Warhol arc in the BCG-subtracted NIRCam images. The {\it HST} and {\it JWST} photometry of the Warhol arc is listed in Table \ref{tab:arc_photometry}.

Our fitting procedure uses {\tt fsps} to construct model spectra of composite stellar populations. The model depends on the following free parameters: the stellar metallicity $\log(Z_*/Z_\odot)$,  nebular ionization parameter $\log(U_{\rm neb})$, $V$-band attenuation due to dust $A_V$, and star-formation rate (SFR) in seven independent temporal bins. The first three bins span the most recent 0--10\,Myr in lookback time (0--2\,Myr, 2--5\,Myr, 5--10\,Myr), and the following four bins are evenly spaced in $\log({\rm time})$ up to the onset of star formation at $z=20$.  A Milky Way dust attenuation curve is assumed from \cite{Cardelli_1989} with $R_V=3.10$. We use a \cite{Salpeter_1955} IMF with the linear slope $\alpha=2.35$ for the mass range 0.08--120\,$M_\odot$. 

The {\tt fsps} code generates a composite model galaxy spectrum including the combined stellar continuum and the nebular emission. We fix the nebular metallicity $\log(Z_{\rm neb}/Z_\odot)$ to be equal to the stellar metallicity, as the {\tt fsps} documentation recommends in order to generate accurate emission-line ratios. 

For each model {\tt fsps} spectrum, we use {\tt sedpy} \citep{johnson_2021_sedpy} to generate synthetic photometry in the {\it HST} and {\it JWST} filters for which we have measured the Warhol arc's photometry. A list of emission-line luminosities for each model is generated by {\tt fsps}, allowing us to include the observed MUSE and LBT emission-line fluxes in our fitting procedure.

The probability density for a given {\tt fsps} model with a set of model parameters $\theta$ is computed using the Gaussian probability density,

\begin{equation}
    p(\theta) \propto {\rm exp} \left[-\frac{1}{2}\sum_{i} \frac{(x_i(\theta)-\mu_i)^2}{\sigma_i^2}+{\rm ln}(\sigma_i^2)  \right]\, ,
\end{equation}

\noindent where $x(\theta)$ is the synthetic photometry or emission-line flux from the {\tt fsps} model for a given set of model parameters $\theta$, $\mu$ is the observed photometry or emission-line flux of the Warhol arc, $\sigma$ is the 1$\sigma$ uncertainty in the observed flux measurement, and the sum is over all the observed {\it HST} and {\it JWST} filters and emission-line fluxes.

We use {\tt emcee} to explore the parameter space using 32 walkers and 10,000 steps. The medians of the posteriors are taken as the best-fit values of each parameter, and the 84\% confidence interval as the 1$\sigma$ uncertainties. The best-fit values from the {\tt fsps} fit, along with the priors placed on each parameter, are shown in Table~\ref{tab:fsps_parameters}. The mean and width of the normal priors are set by the values inferred from the emission-line measurements (see Section \ref{subsec:linefluxes}). The reduced chi-squared statistic $\chi^2_\nu$ for the best-fitting SED model is $\chi^2_\nu=0.80$. The best-fitting {\tt fsps} spectrum and star-formation history (SFH) for the Warhol arc are shown in Figure \ref{fig:bestfit_235}. We calculate the surviving stellar mass from the posteriors on the SFH and find $\log(M_*/M_\odot)=10.46\pm0.05$ (not corrected for magnification).

{To ensure that the choice of prior does not drive the resulting best-fit values, we repeat the {\tt fsps} fitting using Uniform priors for each parameter. The best-fit values and uncertainties do not significantly change when using Uniform priors compared to Normal priors.}  

In order to test the robustness of our SED-fitting method, we generate a synthetic {\tt fsps} spectrum with known values for the SFH, $\log(Z_*/Z_\odot)$, $\log(U)$, and $A_V$. We use {\tt sedpy} to generate synthetic photometry in the {\it HST} and {\it JWST} filters and set the uncertainties such that the S/N in each filter is equal to that of the observed photometry for the Warhol arc. We also generate the synthetic emission-line fluxes for the simulated spectrum and set the uncertainties such that the S/N for each emission line is equal to that of the observed Warhol arc MUSE and LBT emission lines.

Performing the modeling process on the simulated SED, all of the model parameters, including the SFR in every temporal bin, can be recovered within the 84\% confidence interval of the posteriors from the {\tt emcee} sampling. Figure \ref{fig:fsps_sim} shows the input and recovered SED, SFH, and model-parameter posterior probability distributions.

In order to test the effect of the IMF slope on the inferred SFH of the Warhol arc, we repeat the {\tt fsps} SED fitting for a range of values of the linear IMF slope $\alpha$ in the range 0.0--4.0, with a step size of $\alpha=0.10$ (i.e., 41 different values of $\alpha$). The best-fitting SFH for each IMF slope is used to generate the stellar population in our simulations (see Section \ref{sec:simulation}). The best-fitting SFHs for each value of $\alpha$ are shown in Figure  \ref{fig:SFH1} and Figure \ref{fig:SFH2}.

All IMF slopes with $\alpha<3.5$ can produce SED models for the Warhol arc that reproduce the observed photometry and emission lines within the statistical uncertainties ($\chi^2_\nu \approx1.0$). The extremely steep IMF slopes ($\alpha>3.5$) produce SED models that are not consistent with the observed data ($\chi^2_\nu >> 1.0$). This is due to the fact that ionizing radiation from hot, massive stars is necessary to produce the observed strong nebular emission lines, and extremely steep IMF slopes result in very few of these stars. 

\begin{figure}
    \centering
    \includegraphics[width=\linewidth]{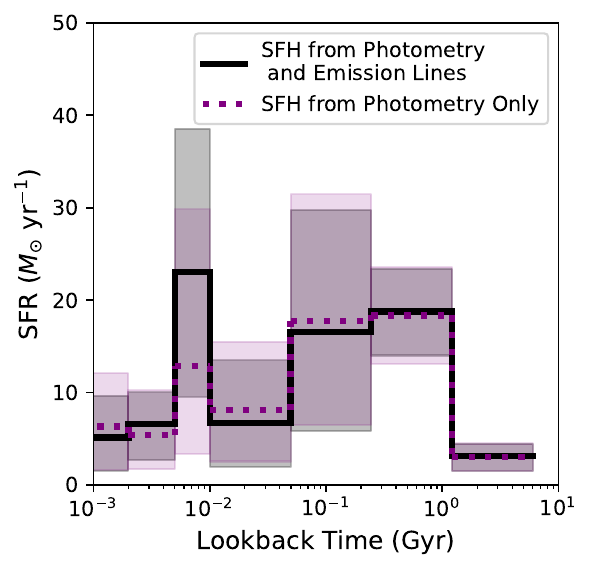}
    \caption{Inferred SFH for the Warhol arc from our {\tt fsps} fitting. The black line shows the SFH inferred using the {\it HST}+{\it JWST} photometry and the MUSE+LBT emission-line measurements as constraints, and the purple line shows the SFH inferred using only the {\it HST}+{\it JWST} photometry. The shaded regions indicate the 1$\sigma$ uncertainties. The SFRs are not corrected for magnification.}
    \label{fig:SFH_comp}
\end{figure}

To test whether the constraints provided by the observed emission lines significantly impact the results of the SED fitting, we repeat the {\tt fsps} modeling without using the emission-line measurements; only the {\it HST} and {\it JWST} photometry is included as a constraint. The best-fitting SFH from the photometry-only version is nearly identical to the SFH from the version including the emission lines in all but one of the temporal bins; the SFR of the burst in the 5--10\,Myr bin is a factor of $\sim2$ smaller for the photometry-only version (see Figure \ref{fig:SFH_comp}).

\begin{table}[ht!]
    \caption{{\tt fsps} SED Modeling Parameters}
    \label{tab:fsps}
    \centering
    \ra{1.3}
    \begin{tabular}{ccc}
    \toprule\toprule
    \textbf{Parameter} & \textbf{Prior [range]} & \textbf{Best-Fit Value$^a$}\\
    \midrule

    $\log(Z_*/Z_\odot)$ & Normal [-2, 0.19] & -0.21 $\pm$ 0.05 \\
    & $\mu=-0.24$, $\sigma=0.08$ &\\

    $\log(U_{\rm neb})$ & Normal [-4, -1] &-3.25 $\pm$ 0.04\\

    & $\mu=-3.30$, $\sigma=0.12$ & \\
    
    $A_V$ & Normal [0, 2] & 0.30 $\pm$ 0.05\\
    & $\mu=0.30$, $\sigma=0.18$& \\

    \bottomrule
    \end{tabular}
    \label{tab:fsps_parameters}
    \footnotesize{$a:$ Best-fit values are for the canonical IMF slope ($\alpha=2.35$) version of the model.}
\end{table}

\section{ Simulation of Microlensing Events}\label{sec:simulation}

To compute the radial and tangential magnification, we use a {\tt GLAFIC} model (v4) \citep{oguri10, kawamataoguriishigaki16, kawamataishigakishimasaku18}, but shift the critical curve to match better the location of the arc's line of symmetry. 

For each IMF slope, we use the inferred SFH for that value of the IMF (see Section \ref{sec:prospector}), as well as the grid spacing, to assign weights to points on the MIST isochrone \citep{Dotter2016,Choi2016}. Integrated over stellar masses, these weights reproduce the input SFH. Integrated over time, these weights reproduce the input IMF.  

For each pixel, we next renormalize all of the weights so that the total flux of the stars in the isochrone matches the pixel's F125W flux density, corrected for magnification. To estimate each pixel's F125W flux density, we use the flux measured from BCG-subtracted images produced by \citet{Shipley_2018}.

The renormalized weights correspond to an expectation value for the number of each star. We next draw from a Poisson distribution to compute the number of stars at each point on the isochrone.  

In the step that follows, we use the {\tt M\_SMiLe} code \citep{Palencia_2024} to approximate the probability distribution of magnification for each star on the isochrone. {\tt M\_SMiLe} computes the probability of potential magnification values given the local shear and convergence, as well as the stellar mass density of the intracluster stars. Modeling the SED of the intracluster light (ICL) using the same method described in Section \ref{sec:prospector}, the inferred intracluster stellar-mass density is $\Sigma_{\rm ICL} = 54 \pm 6$\,$M_{\odot}$\,pc$^{-2}$, which is consistent with the value $\Sigma_{\rm ICL} = 59$\,$M_{\odot}$\,pc$^{-2}$ measured by \citet{Kaurov_2019} for the Warhol arc. 

Adjacent to the critical curve corresponding to a fold caustic, the magnification falls inversely with offset from the critical curve. As listed in Table 2 of \citet{Chen_2019}, pre-{\it JWST} models show a standard deviation of $\sim 20$\% in the normalization of the $1/R$ dependence of the magnification at an offset of 0.06$''$. 
We perform 100 Monte Carlo simulations while drawing randomly from the MCMC chains computed when fitting the arc's SED (thereby incorporating the uncertainties in the SFH, dust attenuation, and stellar metallicity; see Section \ref{sec:sed_fitting}), and also randomly drawing from the 20\% uncertainty in magnification, and from the uncertainty on the mass density of intracluster stars.



\section{Results}\label{sec:results}

\subsection{Stellar Temperatures and Magnitudes}

\subsubsection{Observed temperatures and luminosity requirements}

The temperatures of the nine Warhol stars are in the range $\log(T/\rm K) =3.53$--3.77 ($T=3400$--5800\,K), and two of these cool stars likely have hot B-type binary companions. Assuming a maximum possible magnification of $\mu=10,000$, a star with $\log(T/\rm K) =3.65$ would need to have a minimum luminosity of $\log(L/L_\odot)>3.12$ in order to be detectable in at least two filters given the 5$\sigma$ detection limits (see Table \ref{tab:detection}). With this luminosity constraint, main-sequence stars having temperatures in the range $\log(T/\rm K) =3.53$--3.77 would not be bright enough to be detected. Therefore, the observed stars in the Warhol arc are likely all post-main-sequence stars, either giants or supergiants.

Since giants and supergiants have extremely large stellar radii of $100 < R/R_\odot < 1500$ \citep{Levesque_2009}, and the maximum magnification of a source is inversely proportional to the square root of its radius \cite{Miralda_Escude_1991}, the maximum possible magnifications of giant stars are much lower than they would be for a more compact star. For example, if we assume that the maximum possible magnification of a star with $R=1.0R_\odot$ is $\mu\approx10,000$, a star with $R=100\,R_\odot$ would have a maximum magnification of $\mu\approx1000$. 

Assuming a maximum magnification for giant stars of $\mu=1000$, the faintest source (W8) requires a minimum luminosity of $\log(L/L_\odot)>4.29$ to reproduce the observed fluxes, and the brightest source (W1) requires a minimum luminosity of $\log(L/L_\odot)>5.19$. These luminosity requirements place the stars in the supergiant luminosity class, suggesting that all nine lensed stars are red supergiants originating from massive stars with $M\gtrsim10\,M_\odot$ that have evolved off the main sequence.

{The inferred dust extinctions along the individual lines of sight to each star are listed in Table \ref{tab:best_fits}. The mean extinction across the individual lines of sight to the nine transients is $A_V=2.08\pm0.86$, which is approximately 1.8~mag ($2\sigma$) greater than the the total dust attenuation for the integrated light from the Warhol arc inferred from the observed Balmer decrements ($A_V =0.30\pm0.18$, see Section \ref{subsec:linefluxes}). This discrepancy may indicate the presence of circumstellar dust surrounding the lensed RSGs.}

\subsubsection{Comparison with the simulations}\label{subsubsec:temp_comparison}
\begin{figure}
    \centering
    \includegraphics[width=\linewidth]{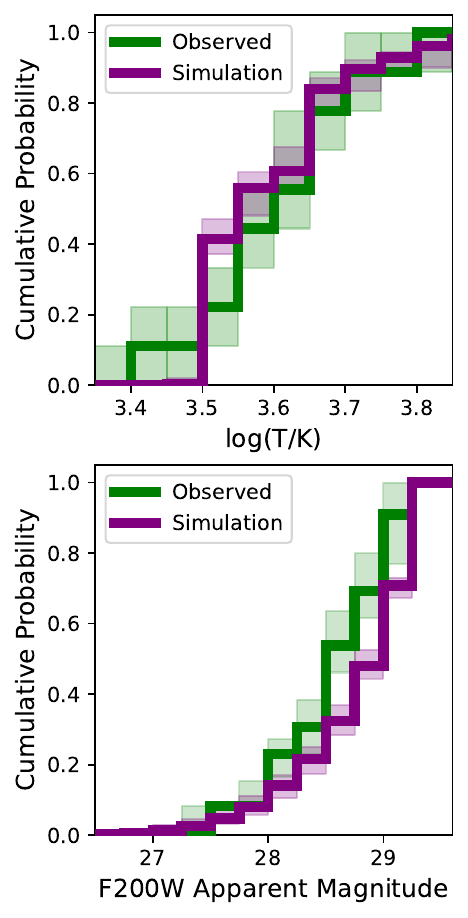}
    \caption{Observed and simulated CDFs for the distribution of apparent magnitudes and effective temperatures of the transients. The simulated CDFs are computed from the $\alpha=2.35$ version of the simulations. The shaded regions show the 1$\sigma$ uncertainty ranges.}
    \label{fig:CDF}
\end{figure}

To compare the simulated stellar temperatures and apparent magnitudes with those of the observed stars, we compute the empirical cumulative distribution functions (CDFs) for the simulated and observed populations. To propagate the uncertainties from the stellar SED fitting and the photometry, we randomly draw temperatures from the MCMC chains and apparent magnitudes from the 150 flux measurements of each transient (see Section \ref{subsec:photometry}). We compute the simulated CDFs for each IMF slope.

To test whether the distributions in temperature and apparent magnitude are consistent between the simulated and observed samples, we perform the two-sample Kolmogorov-Smirnov (K-S) test. The K-S test can be used to infer whether two samples are drawn from the same underlying distribution. Two identical samples would give a K-S statistic equal to zero, and the K-S statistic approaches unity for maximally distinct samples. The significance of the K-S test depends on the size of each sample, and a $p$-value less than 0.05 would indicate that the two distributions are statistically inconsistent. The simulation using the canonical IMF slope ($\alpha=2.35$) produces a distribution of stars that are consistent with the observations in both temperature ($p=0.39$) and F200W apparent magnitude ($p=0.16$). After propagating the uncertainties in both the observations and the simulations, the observed and simulated CDFs are consistent within the 2$\sigma$ uncertainties in all bins (see Figure \ref{fig:CDF}). We repeat the K-S test for all values of the IMF slope used in the simulations, and find that there is no significant dependence on IMF slope. For all tested values of the IMF slope, the K-S test yields $p>0.05$ for both the temperature and magnitude distributions (see Figure \ref{fig:KS}).

\begin{figure}
    \centering
    \includegraphics[width=\linewidth]{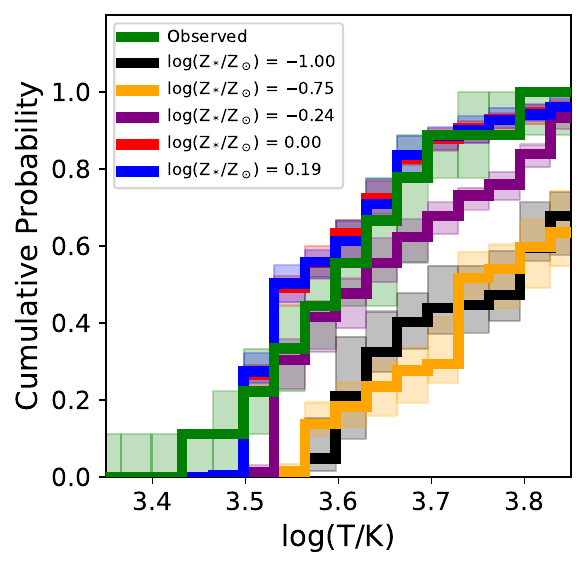}
    \caption{Observed and simulated stellar temperature CDFs for a range of simulated stellar metallicities. The green line shows the CDF for the observed temperatures.}
    \label{fig:metal_temps}
\end{figure}

We test for dependence on stellar metallicity by performing the K-S test using the simulations with the following values of $\log(Z_*/Z_\odot)$: $-1.00$, $-0.75$, $-0.24$, $0.00$, and $0.19$. The stellar metallicity does not significantly affect the simulated distribution of apparent magnitudes, but does affect the simulated stellar temperature distributions. For the version of the simulation which uses $\log(Z_*/Z_\odot)=-0.24$ (i.e., equal to the nebular oxygen abundance we inferred for the Warhol arc from strong emission-line flux ratios), the simulation produces a sample of stars with temperatures that are statistically consistent with the observed stars in the Warhol arc. {The versions of the simulation using higher metallicities ($\log(Z_*/Z_\odot)=0.00$ and $\log(Z_*/Z_\odot)=0.19$) also yield simulated stars whose temperatures are statistically consistent with the observed stars.} For the more metal-poor versions of the simulation ($\log(Z_*/Z_\odot)=-0.75$ and $\log(Z_*/Z_\odot)=-1.00$), the simulated temperature distributions skew significantly hotter compared to the observed temperatures, and the K-S test indicates that the metal-poor simulations are inconsistent with the observed temperatures. Figure \ref{fig:metal_temps} compares the observed and simulated temperature distributions for the tested range of stellar metallicities, and Table \ref{tab:metal_ps} lists the $p$-values from the K-S test.

\begin{table}
    \centering
    \setlength{\tabcolsep}{12pt}
    \ra{1.5}
    \caption{Temperature distribution K-S statistics$^a$}
    \begin{tabular}{ccc}
    \toprule \toprule
    
    Metallicity & KS Statistic & $p$-value\\
    \midrule
       $\log(Z_*/Z_\odot)=-1.00$  &0.54& 0.0058  \\
       $\log(Z_*/Z_\odot)=-0.75$   & 0.60&0.0017\\
       $\log(Z_*/Z_\odot)=-0.24$  &0.30 & 0.32 \\
       $\log(Z_*/Z_\odot)=0.00$  & 0.31 & 0.30\\
       $\log(Z_*/Z_\odot)=0.19$  &0.32 & 0.27\\
       \bottomrule
    \end{tabular}

    \footnotesize{a: K-S statistic and $p$-values for the simulated vs. observed stellar temperatures, over a range of stellar metallicities.}
    
    \label{tab:metal_ps}
\end{table}

\subsection{Detection Rates and IMF Dependence}
\begin{table*}[ht!]
    \caption{Observed vs. Simulated Detection Rates}
    \label{tab:obs_rates}
    \centering
    \ra{1.33}
    \begin{tabular}{cc|cc|cc|cc}
    \toprule \toprule
        & \bf \footnotesize Observed& \multicolumn{2}{c|}{$\alpha=2.35$ (canonical)}   & \multicolumn{2}{c|}{$\alpha=1.00$ (top-heavy)}   & \multicolumn{2}{c}{$\alpha=3.00$ (steep)} \\
    \bf Filter & \bf \footnotesize Detection Rate & \bf Sim Rate & \bf Sim / Obs & \bf Sim Rate & \bf Sim / Obs &\bf Sim Rate & \bf Sim / Obs \\

    \midrule
F200LP & $1.00 \pm 0.71$ & $1.7^{+1.3}_{-0.7}$ &  $1.7 \pm 1.4$ (0.5$\sigma$) & $2.3^{+1.9}_{-1.2}$ &  $2.3 \pm 2.1$ (0.6$\sigma$) & $2.2^{+1.9}_{-0.9}$ &  $2.2 \pm 1.8$ (0.7$\sigma$) \\
F350LP & $2.00 \pm 1.00$ & $1.2^{+1.7}_{-0.7}$ &  $0.6 \pm 0.9$ (-0.4$\sigma$) & $1.9^{+1.5}_{-1.0}$ &  $1.0 \pm 0.9$ (-0.0$\sigma$) & $1.7^{+1.1}_{-0.7}$ &  $0.9 \pm 0.7$ (-0.2$\sigma$)\\ 
F090W & $1.25 \pm 0.56$ & $3.9^{+2.8}_{-1.5}$ &  $3.1 \pm 1.8$ (1.1$\sigma$) & $8.2^{+3.1}_{-2.9}$ &  $6.6 \pm 3.7$ (1.5$\sigma$) & $3.3^{+1.7}_{-1.0}$ &  $2.7 \pm 1.4$ (1.2$\sigma$) \\ 
F115W & $1.50 \pm 0.61$ & $5.4^{+3.5}_{-2.2}$ &  $3.6 \pm 2.1$ (1.2$\sigma$) & $10.9^{+3.8}_{-4.6}$ &  $7.3 \pm 4.3$ (1.5$\sigma$) & $3.8^{+2.4}_{-1.5}$ &  $2.5 \pm 1.4$ (1.1$\sigma$) \\
F150W & $2.50 \pm 0.79$ & $10.4^{+6.9}_{-5.7}$ &  $4.2 \pm 2.6$ (1.2$\sigma$) & $19.0^{+8.3}_{-6.0}$ &  $7.6 \pm 3.4$ (1.9$\sigma$) & $7.2^{+3.6}_{-3.3}$ &  $2.9 \pm 1.6$ (1.2$\sigma$) \\ 
F200W & $3.00 \pm 0.87$ & $23.7^{+12.6}_{-9.2}$ &  $7.9 \pm 3.8$ (1.8$\sigma$) & $37.0^{+16.0}_{-12.0}$ &  $12.3 \pm 5.3$ (2.1$\sigma$) & $15.1^{+9.3}_{-8.0}$ &  $5.0 \pm 3.0$ (1.3$\sigma$) \\ 
F277W & $4.00 \pm 1.00$ & $37.5^{+17.7}_{-19.0}$ &  $9.4 \pm 5.3$ (1.6$\sigma$) & $53.1^{+22.6}_{-16.4}$ &  $13.3 \pm 5.3$ (2.3$\sigma$) & $25.2^{+14.2}_{-12.5}$ &  $6.3 \pm 3.5$ (1.5$\sigma$) \\ 
F356W & $4.00 \pm 1.00$ & $49.0^{+21.7}_{-23.1}$ &  $12.2 \pm 6.5$ (1.7$\sigma$) & $66.7^{+31.1}_{-20.2}$ &  $16.7 \pm 6.5$ (2.4$\sigma$) & $39.3^{+21.0}_{-19.6}$ &  $9.8 \pm 5.5$ (1.6$\sigma$) \\ 
F410M & $3.50 \pm 0.94$ & $18.8^{+11.0}_{-9.8}$ &  $5.4 \pm 3.1$ (1.4$\sigma$) & $26.8^{+13.2}_{-9.0}$ &  $7.7 \pm 3.3$ (2.0$\sigma$) & $14.0^{+6.6}_{-7.4}$ &  $4.0 \pm 2.4$ (1.3$\sigma$) \\ 
F444W & $3.75 \pm 0.97$ & $25.9^{+15.3}_{-11.4}$ &  $6.9 \pm 3.5$ (1.7$\sigma$) & $34.8^{+17.1}_{-10.3}$ &  $9.3 \pm 3.7$ (2.3$\sigma$) & $16.9^{+7.0}_{-8.3}$ &  $4.5 \pm 2.5$ (1.4$\sigma$) \\
\bottomrule
    \end{tabular}
\end{table*}

\begin{figure*}[ht]
    \centering
     \includegraphics[width=\linewidth]{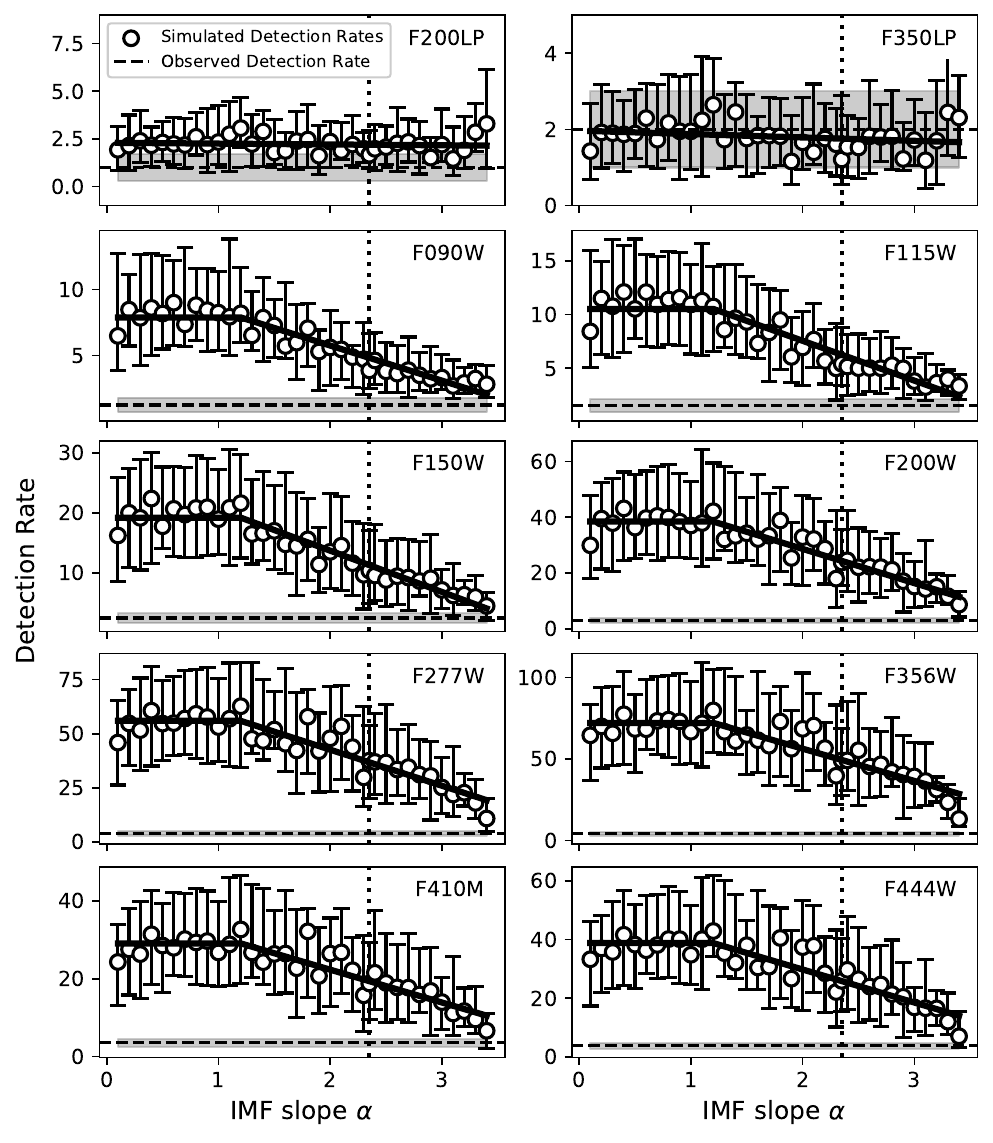}
     \caption{Simulated (open circles with errors) versus observed (dashed line) transient detection rates as a function of the IMF slope for the two {\it HST} flashlights filters (F200LP and F350LP) and all eight {\it JWST} NIRCam filters. The shaded gray region shows the Poisson uncertainty on the observed detection rates. The solid black line gives the piecewise linear fit for detection rate as a function of IMF slope, with the break at $\alpha=1.2$. The dotted vertical line indicates the canonical IMF slope, $\alpha=2.35$.}
     \label{fig:detection_rates}
\end{figure*}

We compare the simulated transient detection rate as a function of IMF slope to the observed transient detection rate in each filter. Table \ref{tab:obs_rates} lists the observed detection rate for each filter, and the simulated detection rate for three values of the IMF slope: $\alpha=2.35$ (canonical), $\alpha=1.0$ (top-heavy), and $\alpha=3.0$ (steep). Figure \ref{fig:detection_rates} shows the simulated detection rates for each IMF slope compared to the observed detection rate in each filter.

For the two {\it HST} {\it flashlights} filters, the simulated detection rates using the canonical IMF slope are consistent with the observed rates within the 1$\sigma$ uncertainties. For all of the {\it JWST} filters, the simulated detection rates using the canonical IMF slope are higher than the observed detection rates by a factor of $\sim 3$--12, but the differences are less than the $2\sigma$ uncertainties in all filters. 

The simulated detection rates for the top-heavy IMF slope are higher than those predicted by the canonical IMF slope in all filters. For the {\it JWST} filters, the top-heavy IMF produces detection rates that are higher than the observed rates by a factor of 6--17, and the discrepancies are $\gtrsim2\sigma$ in the long-wavelength filters. The simulated detection rates using the steep IMF slope ($\alpha=3.00$) are lower than those predicted by the canonical IMF slope for all eight {\it JWST} filters, but still are higher than the observed rates by a factor of 3--10 ($<2\sigma$).

For each filter, we use a least-squares polynomial fitting algorithm to fit a piecewise line to the expected detection rates as a function of IMF slope $\alpha$. All eight NIRCam filters have a strong negative correlation between expected detection rate and $\alpha$ for the range $1.2<\alpha<3.5$, and the relationship flattens for $\alpha<1.2$. The best-fit piecewise lines are shown in Figure \ref{fig:detection_rates}.

\subsubsection{Dependence on Metallicity}

\begin{figure*}
    \centering
    \includegraphics[]{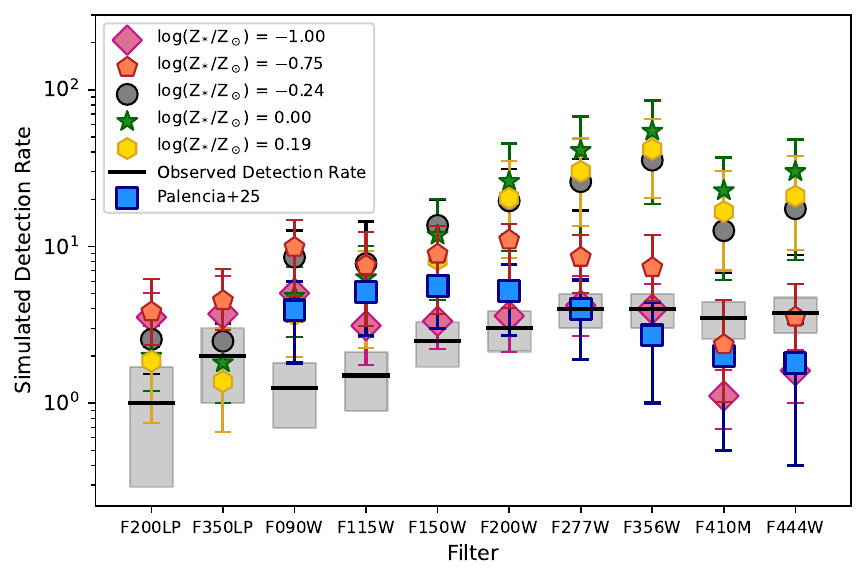}
    \caption{Simulated detection rate as a function of stellar metallicity for the canonical IMF slope $\alpha=2.35$. The black bars show the observed detection rate in each filter, and the gray shaded regions indicate the Poisson uncertainties on the observed detection rates. The blue squares show the simulated detection rate for the Warhol arc from \cite{Palencia_2025} (computed for a fixed detection limit of 29.5\,mag).}
    \label{fig:metal}
\end{figure*}

We compute the expected detection rates for a range of stellar metallicities using the simulations with the canonical IMF slope ($\alpha=2.35$) and the metallicity fixed to the following values of $\log(Z_*/Z_\odot)$: $-1.00$, $-0.75$, $-0.24$, $0.00$, and $0.19$. Since stars at a fixed mass are hotter at lower stellar metallicities, decreasing the stellar metallicity causes the peak of the typical SEDs of the massive, detectable stars in the simulation to shift toward shorter wavelengths. The choice of stellar metallicity significantly impacts the expected detection rates. Figure \ref{fig:metal} shows the simulated detection rate as a function of stellar metallicity. 

For the shortest wavelength filters, the simulated detection rates are higher for the lowest stellar metallicities. For example, the simulated detection rate in the F200LP filter for $\log(Z_*/Z_\odot) = -1.00$ is $3.53^{+1.50}_{-1.24}$, compared to $1.85^{+1.75}_{-1.11}$ for  $\log(Z_*/Z_\odot) = 0.19$. The shift toward higher simulated detection rates for lower metallicities in the short-wavelength filters makes sense, since the typical temperatures of the detectable stars will be hotter, and therefore their SEDs will be bluer on average. 

The impact of the stellar metallicity on the simulated detection rates is most significant for the longest wavelength filters, where the detection rates are lower for higher metallicities owing to the lower temperatures associated with high-metallicity stars. For example, the simulated detection rate in the F356W filter changes by a factor of $\sim10$ for the lowest metallicity compared to the highest metallicity: the detection rate is $3.98^{+1.77}_{-1.26}$ with the metallicity fixed to $\log(Z_*/Z_\odot) =-1.00$, compared to $41.69^{+23.24}_{-21.21}$ for $\log(Z_*/Z_\odot) = 0.19$. 

The simulated detection rates with the stellar metallicity fixed to the most metal-poor values are closest to the observed detection rates in the {\it JWST} filters. For all filters redder than F090W, the simulated detection rates with metallicity fixed to $\log(Z_*/Z_\odot) =-1.00$ or $\log(Z_*/Z_\odot) =-0.75$ are consistent with the observed detection rates within the 1$\sigma$ uncertainties (see Figure \ref{fig:metal}).

This result presents an intriguing tension between the metallicity dependence on the simulated temperature distribution and the metallicity dependence on the simulated detection rates. As discussed in Section \ref{subsubsec:temp_comparison}, the simulated temperature distributions are only consistent with the observed temperature distributions when the choice of stellar metallicity in the simulation is high, $\log(Z_*/Z_\odot)>-0.24$. In contrast, the simulated detection rates (assuming the canonical IMF slope $\alpha=2.35$) in the majority of the filters are only consistent with the observed detection rates when the choice stellar metallicity is low, $\log(Z_*/Z_\odot)<-0.75$. This tension relaxes if we assume a steeper IMF slope of $\alpha=3.0$.


\subsection{Comparison with Complementary Studies}
\cite{Palencia_2025} (hereafter P25) performed a similar but independent detection rate prediction simulation for the Warhol arc. Using a \cite{Kroupa_2001} IMF with $\alpha=2.30$ for stars with $M>1.4$\,$M_\odot$, they estimated simulated detection rates for the Warhol arc in the eight {\it JWST} NIRCam filters. The simulated detection rates from P25 are consistent with our simulated rates within the 1$\sigma$ uncertainties for the short-wavelength NIRCam filters, but the rates from P25 are lower than our simulated rates in the long-wavelength NIRCam filters by a factor of 3--10. 

This discrepancy in the long-wavelength filters likely arises from the metallicity. From their SED-fitting to the photometry of the Warhol arc (split into six different regions), P25 infers significantly lower metallicities ($-1.15<\log(Z_*/Z_\odot)<-0.51$) compared to the value of $\log(Z_*/Z_\odot) = -0.24\pm0.08$ that we measured from the MUSE and LBT emission lines and the value of $\log(Z_*/Z_\odot) = -0.21\pm0.05$ that we inferred from our SED fitting to the photometry and emission-line flux measurements of the Warhol arc. Since lower metallicity stars have hotter temperatures at fixed mass, P25 predicts fewer detectable stars at long wavelengths. As shown in Figure~\ref{fig:metal}, decreasing the metallicity causes the predicted detection rates in long-wavelength filters (F200W--F444W) to significantly decrease by a factor of 3--10, making our predictions consistent within the 1$\sigma$ uncertainties with those of P25.

The difference between the low metallicities inferred by P25's SED-fitting and the higher metallicity that we infer likely arises from the fact that our {\tt fsps} model includes nebular emission while P25's {\tt fsps} model only includes the stellar continuum. To test this, we repeat our SED-fitting procedure described in Section \ref{sec:prospector}, but remove the nebular emission from the {\tt fsps} model. The metallicity inferred from this stellar continuum model is $\log(Z_*/Z_\odot)=-0.66\pm0.12$, which is consistent with the values from P25. Our stellar continuum model is shown in Figure \ref{fig:noneb_fit}.

Other key differences between our analysis and that of P25 include the use of a different lens model for the M0416 cluster (P25 uses the WSLAP+ model of the arc \citep{Diego_2024}), the fact that P25 divides the Warhol arc into six separate regions and masks out intracluster globular clusters and the transient locations when measuring the photometry of the arc, and the fact that P25 uses a delayed-tau model to fit the SFH of the Warhol arc, while we use a flexible SFH prescription which fits for the SFR in seven independent temporal bins.


\cite{Meena_2025} (hereafter M25) performed a similar analysis, predicting the transient event rate for 17 different lensed arcs (including the Warhol arc) that have been observed by the {\it HST} flashlights program. M25 compared the total number of predicted events for all 17 arcs to the total number of observed events, testing two different IMF slopes, $\alpha=2.35$ (canonical) and $\alpha=1.0$ (top-heavy). M25 found that the canonical IMF slope resulted in an underprediction of the number of transient events, while the top-heavy IMF slope resulted in an overprediction of transient events. 

While it is challenging to directly compare our results with those of M25 since that analysis combined 17 arcs and ours focuses solely on the Warhol arc, one key difference is the estimate of the stellar mass density of intracluster stars. M25 uses $\Sigma_{\rm ICL}=30$\,$M_\odot$\,pc$^{-2}$, and the detection rate approximately scales with $\Sigma_{\rm ICL}$. This means that M25's simulation includes a lower number of microlensing stars, leading to a lower predicted event rate compared to our analysis. M25 also infers a lower metallicity from their SED fitting of the {\it HST} photometry of the Warhol arc, $\log(Z_*/Z_\odot)=-0.92$.

\section{Summary and Conclusions}\label{sec:conclusion}
Four deep {\it JWST} NIRCam visits on the M0416 cluster have revealed nine transient events in a single lensed galaxy at $z=0.94$ (the Warhol arc). These detections are likely microlensing transient events of massive stars lying close to the critical curve of the cluster. Using the eight-filter NIRCam SEDs, we constrain the effective temperatures and apparent magnitudes of each lensed star. All nine stars are likely red supergiants with temperatures $T_{\rm eff}\approx4000$\,K, and at least one of these stars has a hot B-type binary companion (Williams et al. 2025, in prep.).

We use {\it HST} and {\it JWST}  photometry of the Warhol arc, as well as emission-line flux measurements from MUSE (VLT), LUCI (LBT), and MOSFIRE (Keck) to fit the galaxy's SED with a stellar population synthesis model. We constrain the SFH of the galaxy, assuming different values for the IMF slope $\alpha$ ranging from $\alpha=0$ to $\alpha=4.0$. All IMF slopes below $\alpha=3.5$ can yield a stellar population synthesis fit that reproduces the observed SED within the 1$\sigma$ uncertainties. 

Using a microlensing simulation, we estimate the expected transient detection rate and the expected temperatures of the detected stars. We repeat the simulation for a range of linear IMF slopes over the range $0<\alpha<4$. When the simulation uses a stellar metallicity $\log(Z_*/Z_\odot)$ equal to the nebular oxygen abundance inferred from the observed emission-line flux ratios in the Warhol arc ($\log(Z_*/Z_\odot=-0.24$), the distributions of expected temperatures are consistent with the observed temperature distribution of the detected stars in the Warhol arc, regardless of the IMF slope. For the simulation version which uses a lower stellar metallicity ($\log(Z_*/Z_\odot)=-0.75 $), the simulated temperature distribution skews significantly hotter than the temperature distribution of the observed stars ($p=0.002$). This result suggests that temperature measurements of microlensing events can be used to constrain the stellar metallicity in galaxies at $z\approx1$.

The expected transient detection rates for the {\it HST} Flashlights observer-frame ultraviolet-optical filters (F200LP and F350LP) are consistent with the observed detection rates in those filters, and the expected rates do not strongly depend on the IMF slope. For all eight {\it JWST}  filters, there is a strong negative correlation between the IMF slope and the expected detection rates over the range $1.2<\alpha<3.5$, but the correlation flattens out for $0<\alpha<1.2$. For the canonical IMF slope $\alpha=2.35$, the simulations tend to overpredict the detection rates compared to the observed rates by a factor of 3--12, but the discrepancy is less than $2\sigma$ in all filters. The top-heavy IMF slope ($\alpha=1.0$) overpredicts the detection rates in the {\it JWST}  filters by a factor of 6--17, and the discrepancies are $\gtrsim2\sigma$ in the long-wavelength filters. 

The simulated detection rate is sensitive to the choice of stellar metallicity. Decreasing the metallicity significantly decreases the simulated detection rate in the long-wavelength filters by up to a factor of $\sim 10$. The simulations with stellar metallicity fixed to $\log(Z_*/Z_\odot)\leq-0.75$ yield expected detection rates that are consistent with the observed detection rates for all filters redder than F090W.  

Intriguingly, there is tension between the simulated detection rates and the simulated temperature distributions in terms of which choice of stellar metallicity yields simulation results that are consistent with the observations. The simulated detection rates are consistent with the observed detection rates only for {\it low} stellar metallicities ($\log(Z_*/Z_\odot)\leq-0.75$), but the simulated stellar temperatures are consistent with the observed stellar temperatures only for {\it high} stellar metallicities ($\log(Z_*/Z_\odot)\geq-0.24$). This tension could arise from the fact that the lens model used in this work does not include substructure, which in principle could modify the high magnification tail for lensed transients thus affecting the predicted apparent magnitudes and detection rates, whereas the color distribution is insensitive to substructure as lensing is inherently achromatic. Substructure may be significant for this arc given that the majority of the microlensed stars lie along the inner edge of the cluster critical curve bisecting the arc \citep{Broadhurst_2025}.

{Another possible explanation for this tension is $\alpha$-element enhancement in the Warhol arc. The stellar atmosphere modeling procedure described in Section \ref{sec:sed_fitting} assumes that the relative abundances of oxygen and iron (O/Fe) are equal to the solar value. However, if $\alpha$-element enhancement is present in the Warhol arc, then the iron abundance in the observed stars may be lower than expected given the inferred oxygen abundance. Since iron abundance plays a significant role in determining stellar temperature, this effect could possibly alleviate the tension between the stellar temperature's and detection rate's dependence on metallicity.}

{The uncertainties associated with stellar models of red supergiant stars could also contribute to the tension. The observed spectra of red supergiants in the local Universe are diverse, and stellar atmosphere models often struggle to simultaneously reproduce the rest-optical and rest-NIR SEDs of these extremely luminous stars \citep[e.g.,][]{Lancon_2007, Davies_2013, Levesque_2018}. Since the {\it JWST} NIRCam short-wavelength filters trace the rest-optical SEDs at $z=0.94$ while the long-wavelength filters trace the rest-NIR, it is possible that stellar modeling discrepancies between these two portions of the observed SEDs could impact the inferred effective temperatures and contribute to the tension. Additionally, if the typical lifetimes of the RSGs in the Warhol arc are shorter than the RSG lifetimes assumed by current models, this discrepancy could explain the high simulated detection rate compared to the observed rate.}

We note the following limitations of our simulations. First, the resolution of the microlensing simulations in the source plane is 2\,nanoarcseconds \citep{Palencia_2025}, which corresponds to $\sim 350\,R_{\odot}$ at $z=0.94$. Second, the magnification simulations consist of subtracting two random draws from the magnification probability distribution function, and do not take into account correlation in magnification with time. Third, our isochrone includes only single stars and not binary systems. A future paper could include a simulation that accounts for correlation in magnification with time and includes binary systems. 

As {\it JWST} continues to target galaxy cluster-scale gravitational lenses with deep NIRCam observations, the number of detected microlensing events in caustic-crossing lensed galaxies will certainly grow. A larger sample of observed microlensing events in lensed galaxies will allow us to infer more robust constraints on the IMF slope and stellar metallicities at $z\approx1$. Future {\it JWST} observations of the MACS\,J0416 cluster would be especially valuable, as these would likely yield additional detections of transient events in the Warhol arc. A larger sample size of highly magnified stars in the Warhol arc would allow for a more statistically robust analysis of the stellar temperature distributions and detection rates.

\section{Acknowledgments}\label{sec:acknowledgments}
This work is based on observations made with the NASA/ESA/CSA {\it James Webb Space Telescope}. The data were obtained from the Mikulski Archive for Space Telescopes (MAST) at the Space Telescope Science Institute, which is operated by the Association of Universities for Research in Astronomy, Inc., under NASA contract NAS 5-03127 for {\it JWST}. These observations are associated with {\it JWST} program 1176. We thank the CANUCS team for sharing their data.

P.L.K. acknowledges U.S. National Science Foundation (NSF) AAG program AST-2308051. This work was support by  NASA/{\it HST} grants GO-15936 and GO-16278 from STScI, which is operated by the Association of Universities for Research in Astronomy, Inc., under NASA contract NAS5-26555. A.V.F. is also grateful for the Christopher R. Redlich Fund and many other donors. Grant JPL-1659411 provided support for some of the ground-based follow-up observations. R.A.W. and S.H.C. acknowledge support from NASA {\it JWST} Interdisciplinary Scientist grants NAG5-12460, NNX14AN10G, and 80NSSC18K0200 from GSFC. A.Z. acknowledges support by grant 2020750 from the United States-Israel Binational Science Foundation (BSF) and grant 2109066 from the U.S. NSF, and by Israel Science Foundation grant  864/23.
 
Some of the data presented herein were obtained at the W. M. Keck
Observatory, which is operated as a scientific partnership among the
California Institute of Technology, the University of California, and
NASA; the observatory was made possible by the generous financial
support of the W. M. Keck Foundation.
We recognize the cultural significance of the Maunakea summit and we are grateful for the opportunity to conduct observations there. 

\bibliography{references}{}
\bibliographystyle{aasjournal}

\break
\appendix

\begin{figure*}[h!]
    \centering
    \includegraphics[width=\linewidth]{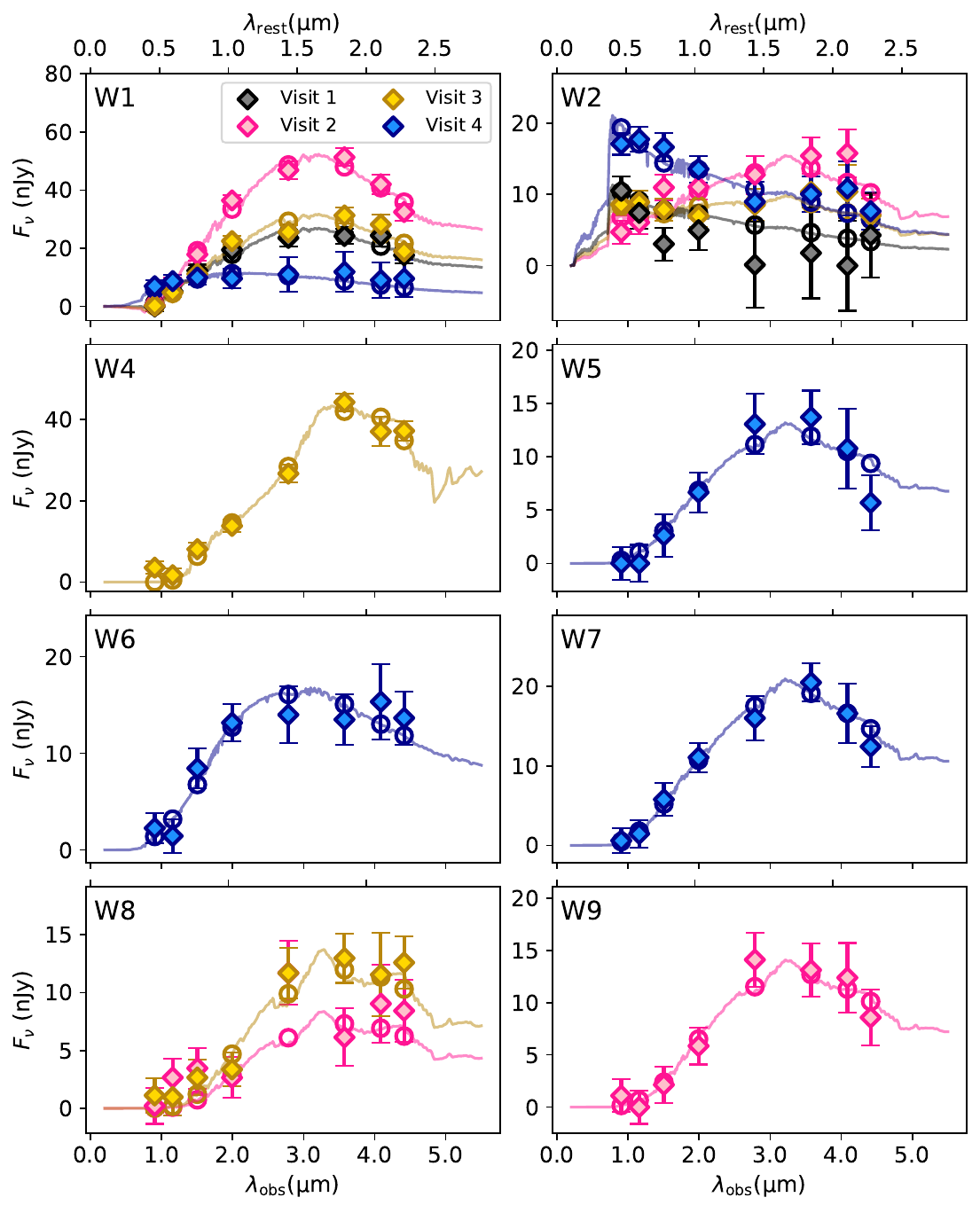}
    \caption{Same as the left panel of Figure \ref{fig:W3_model} but for transients W1, W2, W4, W5, W6, W7, W8, and W9.}
    \label{fig:W1W2W4W5W6W7W8W9_model}
\end{figure*}

\begin{figure*}[h!]
    \centering
    \includegraphics[width=0.9\linewidth]{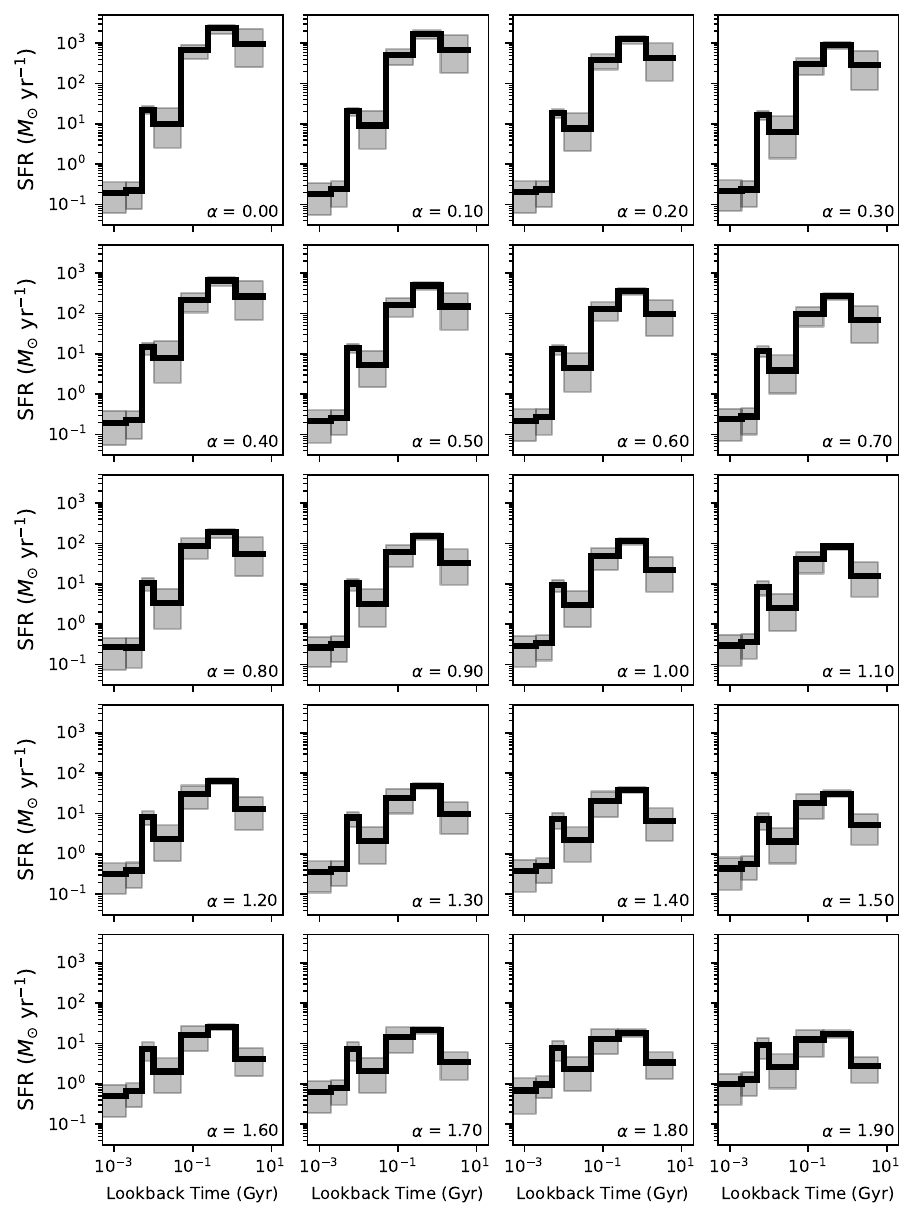}
    \caption{Best-fitting SFH for the Warhol arc from our {\tt fsps} SED modeling, for IMF slopes ranging from $\alpha = 0.0$ to 1.9.}
    \label{fig:SFH1}
\end{figure*}

\begin{figure*}[h!]
    \centering
    \includegraphics[width=0.9\linewidth]{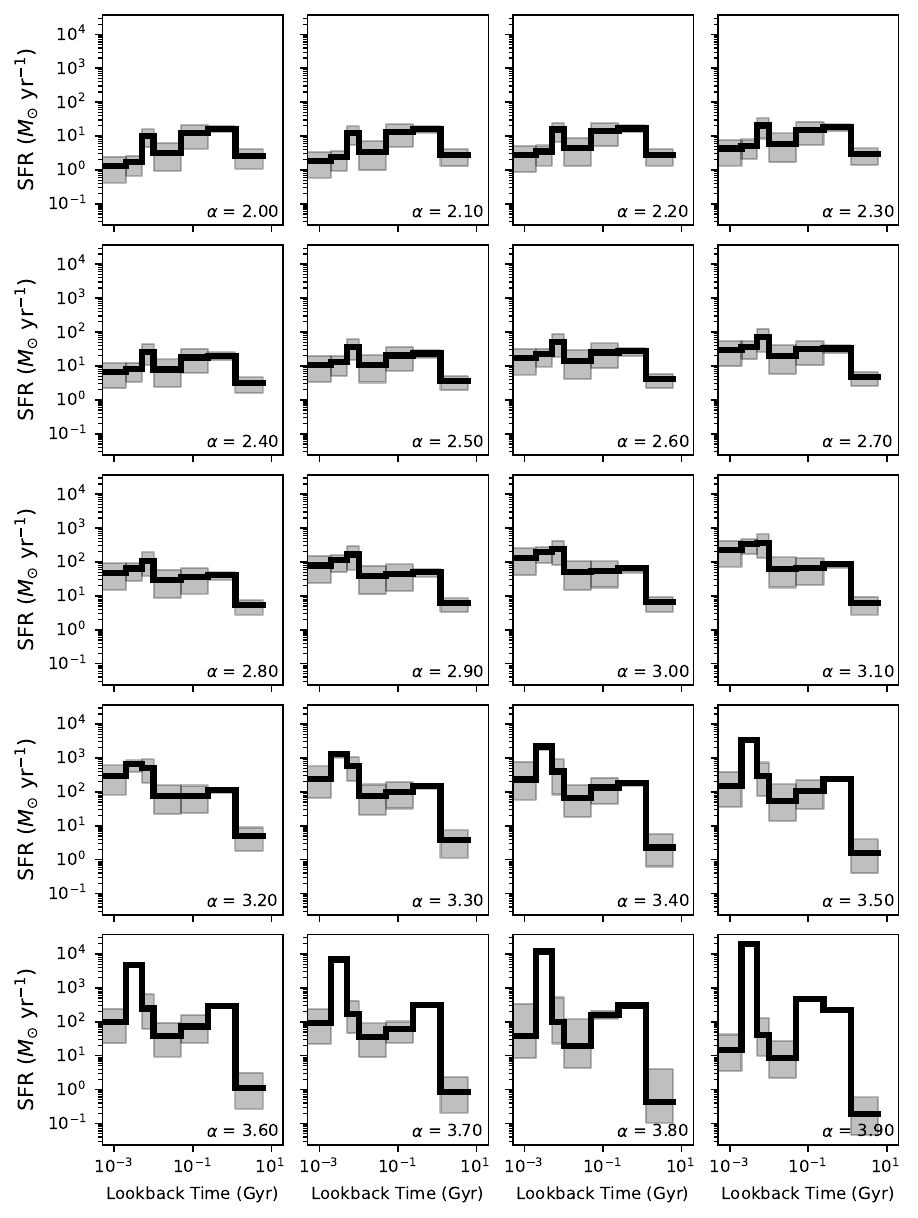}
    \caption{Same as Figure \ref{fig:SFH1}, but for IMF slopes ranging from $\alpha = 2.0$ to  3.9.}
    \label{fig:SFH2}
\end{figure*}

\begin{figure*}
    \includegraphics[width=\linewidth]{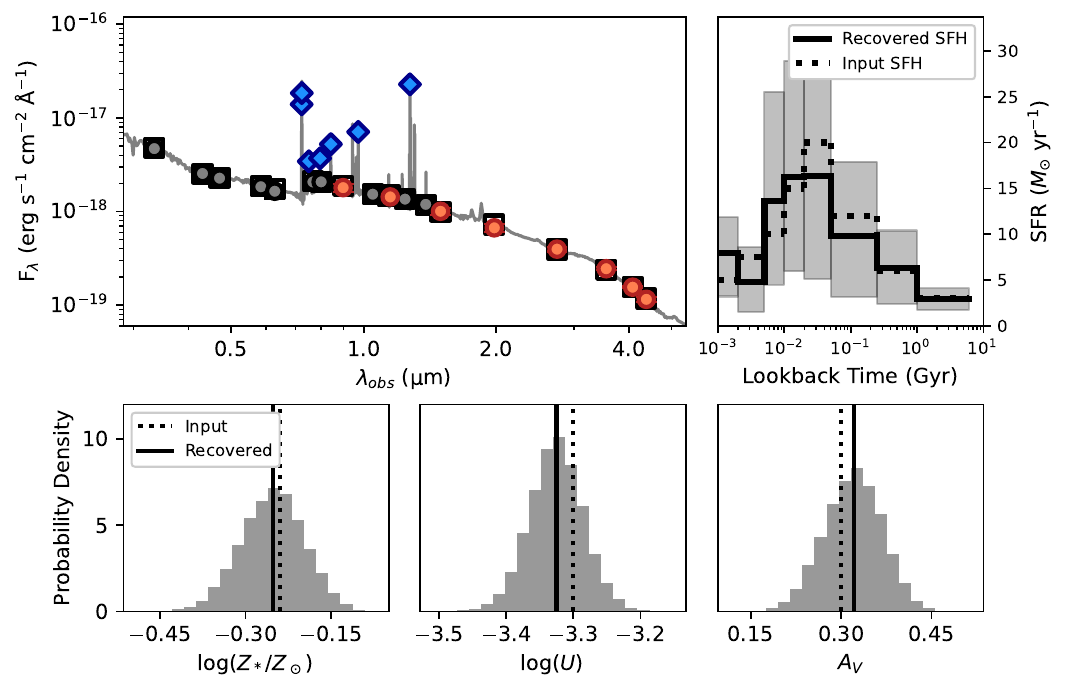}
    \caption{{\it Top}: Synthetic {\it HST} and {\it JWST} photometry of a simulated {\tt fsps} galaxy spectrum with a known input SFH, $\log(Z_*/Z_\odot)$, and $A_V$, along with the recovered spectrum, photometry, and SFH from our SED-fitting procedure. Plotting symbols are the same as in Figure \ref{fig:bestfit_235}. {\it Bottom}: Posterior probability densities from the {\tt emcee} sampling on the simulated data. Our SED-fitting method can recover all model parameters within the 1$\sigma$ uncertainties.}
    \label{fig:fsps_sim}
    
\end{figure*}

\begin{figure*}
    \includegraphics[width=\linewidth]{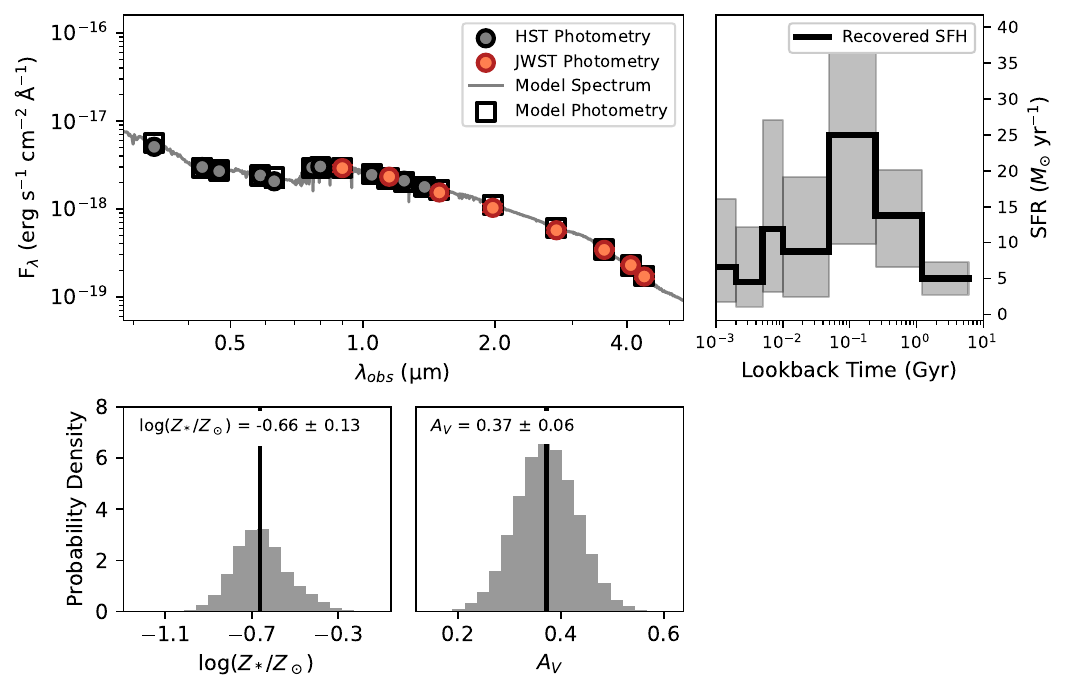}
    \caption{{\it Top}: {\it HST} and {\it JWST} photometry of the Warhol arc, along with the recovered spectrum, photometry, and SFH from the version of our SED-fitting procedure that only includes the stellar continuum. Plotting symbols are the same as in Figure \ref{fig:bestfit_235}. {\it Bottom}: Posterior probability densities from the {\tt emcee} sampling on the simulated data. }
    \label{fig:noneb_fit}
    
\end{figure*}

\begin{figure*}
    \centering
    \includegraphics[width=0.8\linewidth]{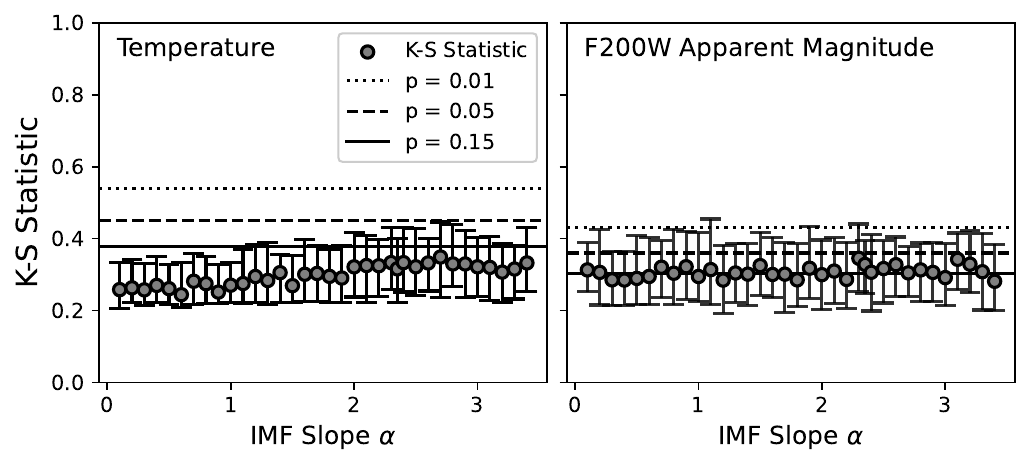}
    \caption{K-S statistic comparing the observed and simulated distributions of stellar temperatures {\it (left)} and F200W apparent magnitudes {\it (right)}. The simulated distributions are not sensitive to the choice of IMF slope.}
    \label{fig:KS}
\end{figure*}

{This Appendix contains the {\it HST} and {\it JWST} photometry of the Warhol arc (Table \ref{tab:arc_photometry}), NIRCam photometry of the nine transient sources in the Warhol arc (Table \ref{tab:photometry}), and the best-fitting values from the stellar atmosphere modeling (Table \ref{tab:best_fits}). Figure \ref{fig:W1W2W4W5W6W7W8W9_model} shows the SEDs of the Warhol transients and the best-fitting model stellar spectra. Figures \ref{fig:SFH1} and \ref{fig:SFH2} display the best-fitting SFHs for the Warhol arc given varying values for the IMF slope. Figure \ref{fig:fsps_sim} illustrates a simulated galaxy SED and the recovered model spectrum from {\tt fsps}. Figure \ref{fig:noneb_fit} gives the stellar continuum-only {\tt fsps} fit to the photometry of the Warhol arc. Figure \ref{fig:KS} shows the K-S statistic comparing the observed vs. simulated stellar temperatures and the observed vs. simulated apparent magnitudes for varying values of the IMF slope.}

\begin{table*}[ht!]
\centering
\ra{1.3}
\setlength{\tabcolsep}{20pt}
\caption{Warhol arc photometry}
\begin{tabular}{ll}
\toprule\toprule

 Filter  & $F_\lambda$ (10$^{-18}$ erg s$^{-1}$ cm$^{-2}$ \r{A}$^{-1})$ \\
\midrule
HST WFC3 UVIS F336w &5.077 $\pm$ 0.233\\
HST ACS WFC F435W & 2.991  $\pm$ 0.0537\\
HST ACS WFC F475W &2.684  $\pm$ 0.186\\
HST ACS WFC F606W &2.401  $\pm$ 0.0873\\
HST ACS WFC F625W &2.059 $\pm$ 0.332\\
HST ACS WFC F775W &2.960  $\pm$ 0.127\\
HST ACS WFC F814W &3.035  $\pm$0.0558\\
HST WFC3 IR F105W &2.440 $\pm$0.0215\\
HST WFC3 IR F110W &2.247 $\pm$ 0.0377\\
HST WFC3 IR F125W &2.091 $\pm$0.0267\\
HST WFC3 IR F140W &1.777 $\pm$0.0244\\
\midrule
JWST NIRCam F090W &2.907 $\pm$0.2739\\
JWST NIRCam F115W &2.322 $\pm$0.2066\\
JWST NIRCam F150W &1.537 $\pm$0.1372\\
JWST NIRCam F200W &1.027 $\pm$ 0.0993\\
JWST NIRCam  F277W &0.571 $\pm$0.03275\\
JWST NIRCam F356W &0.341 $\pm$0.01270\\
JWST NIRCam F410M &0.229 $\pm$0.00456\\
JWST NIRCam F444W &0.169 $\pm$0.01067\\

\bottomrule
\end{tabular}
\label{tab:arc_photometry}

\end{table*}

\begin{table*}[ht]
    \caption{Transient Photometry$^a$}
    \label{tab:photometry}
    \centering
    \footnotesize
    \ra{1.25}
    \begin{tabular}{lcccccccc}
\toprule\toprule
          &\textbf{F090W} &\textbf{F115W}&\textbf{F150W} &\textbf{F200W}&\textbf{F277W}
         &\textbf{F356W}&\textbf{F410M}&\textbf{F444W}   \\
         \midrule
         \textbf{W1}: V1 -- V4 & $<4.62^b$ & 5.09 $\pm$ 1.76 & 12.30 $\pm$ 2.04 & 19.43 $\pm$ 1.93 & 23.56 $\pm$ 2.96 & 24.20 $\pm$ 2.60 & 24.38 $\pm$ 3.88 & 17.57 $\pm$ 2.72 \\
         \textbf{W1}: V2 -- V4 & $<4.32^b$ & 8.13 $\pm$ 1.73 & 17.87 $\pm$ 2.05 & 36.45 $\pm$ 1.73 & 46.87 $\pm$ 3.02 & 51.31 $\pm$ 3.07 & 42.02 $\pm$ 3.39 & 32.46 $\pm$ 2.89 \\
         \textbf{W1}: V3 -- V4 & $<4.68^b$ & 4.79 $\pm$ 1.70 & 11.14 $\pm$ 1.99 & 22.45 $\pm$ 1.70 & 25.56 $\pm$ 2.52 & 31.28 $\pm$ 2.67 & 28.19 $\pm$ 3.46 & 18.95 $\pm$ 2.50\\
        \textbf{W1}: V4 & 6.95 $\pm$ 1.98 & 8.69 $\pm$ 2.26 & 9.92 $\pm$ 2.57 & 9.58 $\pm$ 3.26 & 10.93 $\pm$ 5.91 & 11.88 $\pm$ 6.91 & 9.01 $\pm$ 6.20 & 9.54 $\pm$ 6.37 \\
        \midrule
        \textbf{W2}: V1 &10.51 $\pm$ 2.03 & 7.41 $\pm$ 2.28 & 3.04 $\pm$ 2.31 & 4.98 $\pm$ 2.79 & $<16.7^b$ & $<21.2^b$ &$<12.7^b$ & $<22.5^b$\\
        \textbf{W2}: V2 -- V1 &4.71 $\pm$ 1.59 & 6.07 $\pm$ 1.61 & 10.97 $\pm$ 1.76 & 11.05 $\pm$ 1.78 & 12.76 $\pm$ 2.67 & 15.41 $\pm$ 2.55 & 15.78 $\pm$ 3.34 & 7.17 $\pm$ 2.72  \\
        \textbf{W2}: V3 -- V1 &8.57 $\pm$ 1.51 & 8.91 $\pm$ 1.62 & 7.88 $\pm$ 1.55 & 6.97 $\pm$ 1.50 & 8.54 $\pm$ 2.23 & 10.36 $\pm$ 2.17 & 10.39 $\pm$ 3.73 & 7.05 $\pm$ 2.39 \\
        \textbf{W2}: V4 -- V1 & 17.11 $\pm$ 1.54 & 17.74 $\pm$ 1.75 & 16.64 $\pm$ 2.03 & 13.56 $\pm$ 1.87 & 8.97 $\pm$ 2.84 & 10.05 $\pm$ 2.51 & 10.85 $\pm$ 3.81 & 7.65 $\pm$ 2.59 \\
        \midrule  
        \textbf{W3}: V2 -- V1 & $<4.71^b$ & $<4.83^b$  & 5.24 $\pm$ 1.76 & 8.80 $\pm$ 1.77 & 15.34 $\pm$ 2.74 & 18.06 $\pm$ 2.53 & 10.75 $\pm$ 3.41 & 9.16 $\pm$ 2.77  \\
        \textbf{W3}: V3 -- V1 & $<4.44^b$ & $<4.80^b$ & 4.62 $\pm$ 1.53 & 4.17 $\pm$ 1.48 & 7.83 $\pm$ 2.16 & 9.12 $\pm$ 2.13 & 8.72 $\pm$ 3.55 & 7.78 $\pm$ 2.29 \\
        \midrule
        \textbf{W4}: V3 -- V1& 6.73 $\pm$ 1.52 & 4.23 $\pm$ 1.61 & 8.73 $\pm$ 1.54 & 16.01 $\pm$ 1.50 & 25.95 $\pm$ 2.20 & 43.10 $\pm$ 2.20 & 36.71 $\pm$ 3.62 & 37.05 $\pm$ 2.32\\
        \textbf{W5}: V4 -- V1 & $<4.59^b$ & $<5.22^b$ &  2.62 $\pm$ 2.01 & 6.67 $\pm$ 1.87 & 13.07 $\pm$ 2.83 & 13.72 $\pm$ 2.50 & 10.79 $\pm$ 3.73 & 5.69 $\pm$ 2.57 \\
        \textbf{W6}: V4 -- V1 & 2.26 $\pm$ 1.54 & $<5.28^b$ & 8.45 $\pm$ 2.04 & 13.16 $\pm$ 1.92 & 13.99 $\pm$ 2.97 & 13.49 $\pm$ 2.59 & 15.34 $\pm$ 3.89 & 13.65 $\pm$ 2.75  \\
        \textbf{W7}: V4 -- V1 & $<4.65^b$ & $<5.22^b$&5.78 $\pm$ 2.04 & 11.07 $\pm$ 1.85 & 16.02 $\pm$ 2.76 & 20.50 $\pm$ 2.42 & 16.60 $\pm$ 3.76 & 12.44 $\pm$ 2.55\\
        \midrule
        \textbf{W8}: V2 -- V1 & $<4.47^b$ & $<4.83^b$& 3.47 $\pm$ 1.75 & 2.67 $\pm$ 1.77 & 11.70 $\pm$ 2.75 & 6.14 $\pm$ 2.49 & 9.04 $\pm$ 3.39 & 8.43 $\pm$ 2.69\\
        
        \textbf{W8}: V3 -- V1 & $<4.47^b$ & $<4.83^b$& 2.67 $\pm$ 1.53 & 3.37 $\pm$ 1.48 & 11.69 $\pm$ 2.20 & 12.98 $\pm$ 2.14 & 11.55 $\pm$ 3.60 & 12.59 $\pm$ 2.28\\
        \midrule
        
        \textbf{W9}: V2 -- V1 & $<4.71^b$ & $<4.80^b$&2.14 $\pm$ 1.75 & 5.85 $\pm$ 1.77 & 14.10 $\pm$ 2.57 & 13.11 $\pm$ 2.55 & 12.38 $\pm$ 3.32 & 8.57 $\pm$ 2.67  \\
         \bottomrule
    \end{tabular}
    \footnotesize{$^a$: All flux densities are in nJy.}
    
    \footnotesize{$^b$: 3$\sigma$ upper limit.}
\end{table*}

\begin{table*}[ht!]
\centering
\ra{1.3}
\setlength{\tabcolsep}{16pt}
\caption{Stellar SED Best-Fit Parameters}
\begin{tabular}{llllll}
\toprule\toprule

 & $\log(T /{\rm K})$   & $\log(g / {\rm cm~s}^{-2})$  & $A_V$ (mag)& $R_V$ & $Z_*/Z_{\odot}$ \\
\midrule
W1$^a$  &  $3.68^{+0.03}_{-0.03}$,  $4.02^{+0.30}_{-0.16}$ & $0.81^{+2.35}_{-1.17}$,  $1.72^{+2.30}_{-1.50}$ & $1.37^{+0.58}_{-0.49}$ & $3.89^{+2.46}_{-0.83}$ & $1.11^{+0.58}_{-0.81}$\\
W2$^a$  &  $3.55^{+0.06}_{-0.08}$,  $4.10^{+0.23}_{-0.13}$ & $1.36^{+1.86}_{-1.54}$,  $1.94^{+1.90}_{-1.26}$ & $0.42^{+0.40}_{-0.28}$ & $4.36^{+3.08}_{-1.92}$ & $0.89^{+0.76}_{-0.60}$\\
W3  &  $3.66^{+0.08}_{-0.07}$ & $2.47^{+1.73}_{-2.24}$ & $1.94^{+1.29}_{-1.26}$ & $5.75^{+2.92}_{-3.04}$ & $0.98^{+0.70}_{-0.66}$\\
W4  &  $3.55^{+0.02}_{-0.02}$ & $0.55^{+1.27}_{-1.04}$ & $3.61^{+0.28}_{-0.52}$ & $1.14^{+0.37}_{-0.24}$ & $0.66^{+0.80}_{-0.48}$\\
W5  &  $3.64^{+0.11}_{-0.11}$ & $2.26^{+1.88}_{-2.17}$ & $2.07^{+1.36}_{-1.48}$ & $5.93^{+2.73}_{-3.16}$ & $1.00^{+0.68}_{-0.67}$\\
W6  &  $3.77^{+0.17}_{-0.13}$ & $1.92^{+2.21}_{-2.10}$ & $2.85^{+0.89}_{-1.68}$ & $5.57^{+3.01}_{-3.23}$ & $0.92^{+0.72}_{-0.64}$\\
W7  &  $3.63^{+0.09}_{-0.06}$ & $2.08^{+1.95}_{-2.01}$ & $1.66^{+1.45}_{-1.15}$ & $5.33^{+3.17}_{-3.05}$ & $0.97^{+0.69}_{-0.65}$\\
W8  &  $3.53^{+0.11}_{-0.15}$ & $2.34^{+1.81}_{-2.14}$ & $2.61^{+0.96}_{-1.30}$ & $5.69^{+2.93}_{-3.19}$ & $1.03^{+0.66}_{-0.69}$\\
W9  &  $3.62^{+0.10}_{-0.13}$ & $2.44^{+1.78}_{-2.13}$ & $2.18^{+1.33}_{-1.50}$ & $6.03^{+2.72}_{-2.98}$ & $0.99^{+0.70}_{-0.68}$\\
\bottomrule
\end{tabular}
\label{tab:best_fits}
\footnotesize{$a$:Best-fitting stellar model is a binary system.}
\end{table*}

\end{document}